\documentclass[11pt]{article}
\usepackage[paper=letterpaper,centering,margin=0.8in]{geometry}

\usepackage{amssymb,amsmath,amsthm}
\usepackage{authblk}

\newtheorem{theorem}{Theorem}
\newtheorem{corollary}{Corollary}
\newtheorem{lemma}{Lemma}
\newtheorem{proposition}{Proposition}

\DeclareMathOperator*{\argmax}{argmax}

\usepackage{bm}
\usepackage{xcolor}
\usepackage[ruled,vlined]{algorithm2e}

\usepackage{dsfont}
\usepackage{bbm}
\usepackage{bm}
\usepackage[mathscr]{eucal}
\usepackage{mathtools}
\usepackage[]{color}
\usepackage{tabularx}
\newcolumntype{Y}{>{\centering\arraybackslash}X}

\usepackage{caption}
\usepackage{subcaption}
\usepackage{mwe}
\usepackage{tikz}
\usetikzlibrary{positioning}
\usepackage[ruled]{algorithm2e}
\usepackage{natbib}
 \bibpunct[, ]{(}{)}{,}{a}{}{,}%
\usepackage{enumerate}

\newcommand{\df}{\,\mathrm{d}}

\def\c{{\bm c}}
\def\p{{\bm p}}
\def\x{{\bm x}}

\def\bbeta{{\bm \beta}}

\def\R{\mathbb{R}}
\def\O{{\cal O}}

\def\={\!=\!}

\DeclareMathOperator{\e}{e}

\usepackage[normalem]{ulem}

\newcommand{\Beginproof}{\begin{proof}}
\newcommand{\Endproof}{\end{proof}}

\begin{document}

\title{Perturbed Pricing}

\author{Neil Walton}
\author{Yuqing Zhang}
\affil{University of Manchester\\
\small{\texttt{\{neil.walton, yuqing.zhang \}@manchester.ac.uk}}}

\maketitle

\abstract{
We propose a simple randomized rule for the optimization of prices in revenue management with contextual information. It is known that the certainty equivalent pricing rule, albeit popular, is sub-optimal.
We show that, by allowing a small amount of randomization around these certainty equivalent prices, the benefits of optimal pricing and low regret are achievable.
}

\maketitle

\section{Introduction}\label{section_Perturbed-intro}
When a company sells a new product, the objective is to select prices that maximize revenue.
However, there is often little information about demand, so over time the company must choose prices that maximize long-run revenue and efficiently estimate the distribution of demand.
Moreover, when diverse products are sold online, the demand, and thus prices, depend on the context in which the items are sold, such as who is viewing and what search criteria they used.

A widely-used pricing strategy is the certainty equivalent pricing rule. Here, the manager statistically estimates the demand over the set of prices (and contexts) and then selects the revenue optimal price for this estimated demand, i.e., prices are chosen as if statistical estimates are the true underlying parameters.
This approach is very appealing as it cleanly separates the statistical objective of estimating demand from the managerial objective of maximizing revenue.
However, it is well-known in both statistics and management science literature that a certainty equivalent rule may not explore prices with sufficient variability. This leads to inconsistent parameter estimation and thus sub-optimal revenue. 

Overcoming this difficulty is an important research challenge for which numerous different methods have been proposed. 
We review these shortly, but, to summarize, there are two principal approaches: to modify the statistical objective, or to modify the revenue objective. For instance, if the statistical objective is a maximum likelihood objective,
then likelihood function can be modified via regularization to induce consistency. 
For the revenue objective, confidence intervals and taboo regions can be used to restrict the set of available prices and thus ensure exploration.
These modifications involve additional bespoke calculations that couple the statistical and revenue objectives.

In this article, we advocate a different approach: pricing according to a certainty equivalent rule with a decreasingly small random perturbation. Estimation can then be conducted according to a standard maximum likelihood estimation. We call this \emph{perturbed certainty equivalent pricing}, or \emph{perturbed pricing}, for short.
From a managerial perspective, the key advantage is its simplicity---we perturb the data not the optimization. The statistical objective and revenue objective remain unchanged and can be treated separately. Thus parameters can be estimated using conventional statistical tools, and prices can be chosen with standard revenue optimization techniques. If the magnitude of the perturbation is chosen well, our results prove that perturbed pricing performs comparably with the best pricing strategies.

\subsection{Overview}
We present a brief summary of our model, pricing strategy, results and contributions. A formal description and mathematical results are given in Sections~\ref{section_Perturbed-setup} and \ref{section_Perturbed-result}, respectively.

\smallskip

\noindent \textbf{Model.} 
We let $r(\bm{p}, \bm{c} \,; \bm{\beta}_0)$ be the revenue for a set of products at prices $\bm{p}$ under context $\bm{c}$ given parameters $\bm{\beta}_0$. The \emph{revenue objective} is to find the revenue optimal price for each context:
\[
\bm{p}^\star(\bm{c}) \in \argmax_{\bm{p}}\;\;  r(\bm{p}, \bm{c} \,; \bm{\beta}_0)\,.
\]
The parameters $\bm{\beta}_0$ are unknown; however, these parameters can be inferred.  In particular, we receive a demand response $y$ to the price-context input $\bm{x}=(\bm{p},\bm{c})$ as a generalized linear model:
\[
y = \mu\left(\bm{\beta}_{0}^{\top}\bm{x}\right) + \varepsilon \,.
\]
Here $\mu$ is an increasing function and $\varepsilon$ has mean zero. The response $y$ can be interpreted as the number (or volume) of items sold given the prices and context. Given data  $((\bm{x}_s, y_s) : s=1,...,t)$, the \emph{statistical objective} is a maximum likelihood optimization:
\begin{equation}\label{eq_Perturbed-MLE}
\bm{\hat \beta}_t \in \argmax_{\bm \beta}\;\; \sum_{s=1}^t 
y_s \bm \beta^\top \bm{x}_s - m(\bm \beta^\top \bm{x}_s) \,,
\end{equation}
where ${m}'(z) = \mu(z)$.

\smallskip
\noindent \textbf{Pricing.} Given an estimate $\bm{\hat{\beta}}$, the certainty equivalent price is 
\begin{equation}\label{eq_Perturbed-CEP}
\bm{p}_{ce}(\bm{c} \,; \hat{\bm{\beta}})\in \argmax_{\bm{p}  }\;\;  r(\bm{p}, \bm{c} \,; \hat{\bm{\beta}})\, .
\end{equation}
Notice $\bm{p}^\star (\bm{c}) = \bm{p}_{ce}(\bm{c} \,; \bm{\beta}_0)$.
For the maximum likelihood estimate at time $t-1$, $\bm{\hat \beta}_{t-1}$, and new context, $\bm{c}_t$, the {perturbed certainty equivalent price} is
\begin{equation}\label{eq_Perturbed-PCEP}
	\bm{p}_t  = \bm{p}_{ce} (\bm{c}_t \,; \bm{\hat \beta}_{t-1}) + \alpha_t \bm{u}_t \,,
\end{equation}
where $\bm{u}_t$ is an independent, bounded, mean zero random variable and $\alpha_t$ is a positive real number. The random variables $\bm{u}_t$ and $\alpha_t$ can be selected in advance. For instance, we may choose $\bm{u}_t \sim \text{Uniform}([-1,1]^d)$ or $\bm{u}_t\sim \text{Uniform}(\{\pm \bm e_i: i=1,...d\})$ where $\bm e_i$ is the $i$-th unit vector. We will advocate taking $\alpha_t = t^{-\frac{1}{4}}$.

\smallskip
\noindent \textbf{Results.} The regret measures the performance of the perturbed pricing strategy compared to the revenue optimal strategy:
\[
\mathcal{R}\!g(T) = \sum_{t=1}^T r(\bm{p}^\star(\bm{c}_t),\bm{c}_t \,; \bm{\beta}_0) - r(\bm{p}_t,\bm{c}_t \,; \bm{\beta}_0)\, .
\]
It is known that $\mathcal{R}\!g(T) = \Omega(\sqrt{T})$ for any asymptotically consistent pricing policy. In Theorem \ref{theorem_Perturbed-regret-bound} we prove that for $\alpha_t = t^{-\frac{1}{4}}$ the regret of the perturbed pricing satisfies
\[
\mathcal{R}\!g(T) = \O \left(\sqrt{T} \log (T)\right)\,.
\]

\subsection{Contributions.} 
In Theorem \ref{theorem_Perturbed-regret-bound} we show that the convergence of the perturbed certainty equivalent pricing is optimal up to logarithmic factors. 
The speed of convergence is competitive with the best existing schemes. 
However, as we have already discussed, the main contribution is the simplicity of the policy. 
Current schemes often require additional matrix inversions, eigenvector calculations, or introduce non-convex constraints to the pricing optimization. 
Furthermore the scheme is flexible in leveraging contextual information, which is an important best-practice in many online market places and recommendation systems. 

In forming perturbed certainty equivalent prices, the manager can estimate parameters as a standard maximum likelihood estimator \eqref{eq_Perturbed-MLE}, and can price according the revenue maximization objective \eqref{eq_Perturbed-CEP}. The only change is to introduce a small amount of randomization \eqref{eq_Perturbed-PCEP}. This is appealing as the statistical optimization and revenue optimization remain unchanged and the perturbation of prices is simple, intuitive, and requires negligible computational overhead.

In addition to this managerial insight, there are a number of mathematical contributions in this paper.
The eigenvalue lower bound used in Proposition \ref{proposition_Perturbed-lambdamin-bound} is new and also crucial to our analysis of design matrices. 
We build on the work of \cite{LaiWei1982} to clarify results on the rate of convergence and strong consistency of generalized linear models in Proposition \ref{proposition_Perturbed-Beta-bound}. 
Further, as we will review, much of the current revenue maximization literature on pricing applies to revenue optimization of a single item in a non-contextual setting. To this end, we include the important generalizations of contextual information and the sale of multiple items.

\section{Related Work}\label{section_Perturbed-review}
In this section, we provide a brief review of revenue management with unknown demand, contextual multi-arm bandits, feature-based pricing, and strong consistency of generalized linear models.
\paragraph{Revenue Management with Unknown Demand.}
Optimizing revenue over prices is a central theme in revenue management.
Texts such as \cite{PhillipsRobert2005} and  \cite{TalluriRyzin2005} provide an excellent overview.  In classical works, the demand for each price is known to the decision maker. However both in practice and in more recent academic literature, the demand function must be estimated.
This is clearly articulated in the seminal work of \cite{BesbesZeevi2009}. 
Here the classical revenue management problem of \cite{gallego1994optimal} is recast as a statistical learning problem and shortfall in revenue induced by statistical estimation is analyzed.
Subsequently there have been a variety of studies jointly applying optimization and estimation techniques in revenue management, see \cite{DenBoer2015} for a survey of this literature.

We consider a parametric statistical model. 
The distribution of demand is fixed with unknown parameters. 
Here a popular policy is the certainty equivalent pricing rule. 
It was first introduced by \citet{AndersonTaylor1976} for linear models in econometrics. However, their greedy iterated least squares algorithm is known to be sub-optimal;
see \cite{LaiRobbins1982} for a counter-example and \cite{DenBoer2013} for a counter-example in the context of revenue management. Subsequent works have developed mechanisms to overcome this issue.
\cite{BroderRusmevichientong2012} introduced a maximum-likelihood model with a single unknown parameter. 
They achieved an upper bound with a $\O(\sqrt{T})$ regret in the “well-separated” case, where all prices are informative.
\citet{DenBoerZwart2014S} proposed a controlled variance pricing policy, in which they created \textit{taboo} intervals around the average of previously chosen prices. 
They obtained an asymptotic upper-bound on $T$-period regret of $\O\left(T^{1/2+\delta}\right)$ for arbitrarily small $\delta>0$.
\cite{KeskinZeevi2014} considered the expected demand as a linear function of price and provided general sufficient conditions for greedy iterated least squares.
In particular, their constrained iterated least squares algorithm follows a similarly appealing rationale to our perturbed certain equivalent price. However, the algorithm requires the notion of a time-averaged price which may not be obtainable in a contextual setting.
Both controlled variance pricing policy and, under appropriate assumptions, constrained iterated least squared, can ensure sufficient price dispersion for low regret outcomes.
The perturbed pricing policy considered in this paper was first proposed for the pricing of a single product in \cite{lobo2003pricing}.
The policy is analyzed via simulation (without supporting theoretical analysis) and they noted that randomization can significantly improves result. This work substantially generalizes their setting and is the first paper to provide a theoretical basis for their positive findings.

The revenue management literature discussed so far does not incorporate contextual information on the customer, the query and the product. To this end, we first discuss the contextual multi-arm bandit problem and then feature-based pricing.


\paragraph{Contextual Multi-arm Bandits.}
A multi-arm bandit problem is a broad class of sequential decision making problems where an algorithm must jointly estimate and optimize rewards. For an overview on multi-armed bandit problems, see \citet{BubeckBianchi2012} and \cite{LattimorSzepesvsrie2016}.
Our work is related to literature on contextual multi-armed bandits. Here the algorithm receives additional information which can be used to inform the algorithm's decision.  
This has been applied to many problems, such as clinical trials \citep{Woodroofe1979} and online personalized recommendation systems \citep{Li2010}. 

\citet{Auer2002a} studied multi-armed bandits, where actions are selected from a set of finite features and the expected reward is linear in these features.
Following from \citet{Auer2002a}, algorithms based on confidence ellipsoids were described in 
\citet{Dani2008}, \citet{RusmevichientongTsitsiklis2010}, \citet{AbbasiYadkori2011} and \citet{Chu2011}.
\citet{AbbasiYadkori2011} proved an upper-bound of $\mathcal{O}(d\sqrt{T})$ regret after $T$ time periods with $d$-dimensional feature vectors.
Generalizing previous work on this linear stochastic bandit problem, \cite{Filippi2010} introduced a generalized linear model with the upper confidence bound algorithm and achieved $\widetilde{{\O}}(d\sqrt{T})$ regret.
\citet{Li2017} improved the work of \cite{Filippi2010} by a $\sqrt{d}$ factor.

\paragraph{Feature-based Pricing.}
Following from the above work on contextual bandits,
there is a growing literature on dynamic pricing with features (or covariates). 
The feature information	may help the decision maker to improve estimation and segment distinct market places.
\citet{Amin2014} studied a linear model, where features are stochastically drawn from an unknown i.i.d.\@ distribution.
They proposed a pricing strategy based on stochastic gradient descent, which achieved sub-linear regret $\widetilde{{\O}}(T^{2/3})$.
\citet{Cohen2016} considered a problem similar to \citet{Amin2014}. 
They assumed that the feature vectors are adversarially selected and introduced an ellipsoid-based algorithm, which obtained regret of $\O(d^2 \log(T/d))$.
\citet{Qiang2016} assumed the demand follows a linear function of the prices and covariates, and applied a myopic policy based on least-square estimations  which achieved a regret of $\O(\log(T))$.
\cite{JavanmardNazerzadeh2019} considered a structured high-dimensional model with binary choices and proposed a regularized maximum-likelihood policy which achieves regret $\O(s_0 \log(d) \log(T))$.
These prior models achieve a logarithmic rather than square root regret because demand feedback is a deterministic function of some unknown parameter. See \cite{kleinberg2003value} for an early discussion on this distinction and lower-bounds in both settings. 
\cite{BanKeskin2019} were the first to introduce random feature-dependent price sensitivity and achieved the expected regret of $\O(s\sqrt{T} \log(T))$ and $\O(s \sqrt{T} (\log(d)+\log(T)))$ in linear and generalized linear demand models.
\cite{chen2015statistical} considered statistical learning and generalization of feature-based pricing.

\paragraph{Maximum Quasi-likelihood Estimation for GLMs.}
We model demand with a generalized linear model (GLM).
\cite{McCullaghNelder1989} is a classical reference.
 GLMs are widely used for ratings systems and pricing decisions, see \cite{OhlssonJohansen2010}.
In the GLM framework, a commonly used technique to estimate the parameters is maximum likelihood estimation. 
\citet{Wedderburn1974} proposed maximum quasi-likelihood estimation, as extension of likelihood estimations.

Following our discussion on the counter-example of \cite{LaiRobbins1982}, the consistency of parameter estimators is of great concern.
To deal with this problem, conditions are usually imposed, see \citet{Lai2003} for a more detailed discussion.
When the regression model is linear, \citet{LaiWei1982} proved the strong consistency of estimation when the ratio of the minimum eigenvalue to the logarithm of the maximum eigenvalue goes to infinity.
For GLMs with canonical link functions, \citet{ChenHY1999} derived similar strong consistency results under similar conditions to \citet{LaiWei1982}.
For GLMs with general link functions, \citet{Chang1999} obtained strong consistency via a last-time variable based on a sequence of martingale differences and an additional assumption. 
\cite{denBoerZwart2014M} reported on mathematical errors in the works of \cite{Chang1999} and \cite{ChenHY1999}. Consequently some alternative derivations have been developed by \cite{denBoerZwart2014M} and \cite{Li2017}. 
Based on these, we develop our own strong consistency result in this paper. Also, as discussed, bounds on the design matrix of our GLM are required for convergence. Given the Schur Decomposition and Ostroswki's Theorem (see \cite{horn2012matrix}), we develop a new eigenvalue bound, Proposition 3, that when combined with covariance estimation results from \cite{Vershynin2018} enables us to incorporate random contextual information. 

\section{Problem Formulation}\label{section_Perturbed-setup}

We formally describe the revenue objective, the statistical objective, and the perturbed pricing policy along with assumptions required for our theoretical analysis.

\subsection{Model and assumptions}
\emph{Revenue Optimization.} We choose prices $\bm{p} \in \mathcal{P}$.
Here $\mathcal{P}$ is a bounded open convex subset of $\R^{m}$.
When choosing a price, we receive a context $\bm{c}\in \mathcal{C}$.
This summarizes information about the item (or items) sold and the buyer (or buyers).
Here, $ \mathcal{C} $ is a bounded subset of $\R^{d-m}$, with $d \geq m$.
We let $\bm{x} := (\bm{p}, \c) \in \mathcal{X}$, where $\mathcal{X} := \mathcal{P} \times \mathcal{C} \subseteq \R^{d}$. 

We receive rewards for different prices chosen for different contexts over time. 
Further, there are unknown parameters $\bm{\beta}_{0} \in  \mathcal{B}\subset  \R^{d}$ that influence these rewards.
In particular, we let $r\left(\bm{p}, \bm{c}\,; \bm{\beta}_0\right)$ be the expected real-valued reward for prices $\bm{p}$ under context $\c$ given parameters $\bm{\beta}_{0}$. 
Since $\bm{x}= (\bm{p}, \c)$, we also define $r\left(\bm{x}\,; \bm{\beta}_{0}\right) := r\left(\bm{p}, \bm{c}\,; \bm{\beta}_0\right)$.
We assume that $(\bm{x}, \bm{\beta}) \mapsto  r(\bm{x}\,; \bm{\beta})$ is a twice continuous differentiable function.
Given the context $\bm{c}\in \mathcal{C}$, an objective is to choose prices that maximize reward:
\begin{equation}\label{eq_Perturbed-maxR}
	\bm{p}^{\star}(\c) \in \mathop{\arg\! \max}_{\bm{p} \in \mathcal{P}} \;\; r\left(\bm{p}, \bm{c}\,; \bm{\beta}_0\right).
\end{equation}
The solution $\bm{p}^\star(\c)$ is the {optimal price} for context $\bm c$. 
Given $\bm{\beta}_0$, we assume there is a unique  optimal price $\bm{p}^\star(\c)$ for each $\bm{c}\in\mathcal C$. We place one of two assumptions on the reward function and the set of contexts,
\begin{enumerate}
\renewcommand{\theenumi}{A1\alph{enumi})}
	\item{ The set $\mathcal C$ is finite and the Hessian of $\bm{p}\mapsto  r(\bm{p}, \bm{c}\,; \bm{\beta}_0)$ is positive definite at $\bm{p}^\star(\c)$ for each $\bm{c}\in\mathcal C$.
	}
	\label{ass_Perturbed-A_finite}
\renewcommand{\theenumi}{A1\alph{enumi})}
\setcounter{enumi}{1}
	\item{The set $\mathcal C$ is convex and $\bm{p}\mapsto  r(\bm{p},\bm{c}\,; \bm{\beta}_0)$ is $\alpha$-strongly concave for some $\alpha >0$.
	}
	\label{ass_Perturbed-A_convex}
\end{enumerate}

\medskip
\noindent \emph{Learning Model.} The parameter $\bm{\beta}_{0}$ is unknown. However, we may learn it through statistical estimation.
In particular, we receive responses that are a generalized linear model with parameter $\bm{\beta}_{0}$.
Given $\bm{x} \in \mathcal{X}$, we receive a
response $y$ which is a real-valued\footnote{The analysis given in this paper will likely follow for $y$ in $\mathbb R^n$, but for simplicity we assume $y$ is in $\mathbb R$.} random variable such that
\begin{equation*}
	y = \mu\left(\bm{\beta}_{0}^{\top}\bm{x}\right) + \varepsilon \,.
\end{equation*}
Here $\varepsilon $ is a bounded random variable with mean zero. 
Further, $\mu: \R \to \R$, which is called the
\emph{link function}, is a strictly increasing, continuously
differentiable, Lipschitz function. 

Given data $\left(\bm{x}_{s}, y_{s}\right)$ for $s=1, \ldots, t$, we must estimate unknown parameters $\bm{\beta}_{0}$. 
We let $\hat{\bm{\beta}}_{t}$ denote our estimate of $\bm{\beta}_{0}$.
A popular method for estimating $\bm{\beta}_{0}$ is maximum (quasi-)likelihood estimation. 
Here, maximum quasi-likelihood estimators $\hat{\bm{\beta}}_{t}$ are the solutions to the equations
\begin{equation}\label{eq_Perturbed-MQLE_beta}
    \sum_{s=1}^{t}
    \bm{x}_{s}
    \left(y_{s}-{\mu\left(\hat{\bm{\beta}}_{t}^{\top}\bm{x}_{s}\right)} \right)
    =\bm{0} \, .
\end{equation}
When it exists, the solution to this equation is unique (since $\mu$ is strictly increasing). 
When the distribution of $y_{s}$ given $\bm{x}_{s}$, for $s=1,...,t$, are independent and each belongs to the family of exponential distributions with mean $\mu\big(\hat{\bm{\beta}}_{0}^{\top} \bm{x}_s\big)$, 
then \eqref{eq_Perturbed-MQLE_beta} is the condition on $\bm{\beta}$ for maximizing the log-likelihood. 
In this case, $\hat{\bm{\beta}}_t$ is the maximum likelihood estimator.
However in our case, we don't assume $\left(y_{t}\mid\bm{x}_{t}\right)$ is from an exponential family, so instead, as in \cite{Wedderburn1974}, we refer to $\hat{\bm{\beta}}_{t}$ as maximum quasi-likelihood estimators. 

Typically $\hat{\bm{\beta}}_{t}$ can be found with standard software packages using Newton methods such as Iteratively Reweighted Least Squares. 
For better time complexity, in the case of linear regression, the Sherman--Morrison formula can be applied to yield an online algorithm with $\O(t d^2)$ complexity.

A sequence of estimators $\hat{\bm \beta}$, 
is said to be \emph{strongly consistent} if, as $t \to \infty$ with probability $1$,
\begin{equation*}
\hat{\bm \beta}_t \rightarrow \bm{\beta}_0 \, .
\end{equation*}
For adaptive designs it is often possible to prove even stronger results, specifically that with probability $1$,
\begin{equation}\label{eq_Perturbed-beta_strong}
\left\| \hat{ \bm \beta}_t - \bm \beta_0 \right\|^2 
=
O \left( 
\frac{\log (t)}{\lambda_{\min}(t)} 
\right)  \, .
\end{equation}
where $\lambda_{\min} (t)$ is the minimum eigenvalues of the design matrix $\sum_{s=1}^{t}\bm{{x}}_s \bm{{x}}_s ^{\top}$ and $\left\| \cdot \right\|$ denotes the Euclidean norm.

Our main result, Theorem \ref{theorem_Perturbed-regret-bound}, holds for any generalized linear model such that \eqref{eq_Perturbed-beta_strong} holds. 
We prove, in Proposition \ref{proposition_Perturbed-Beta-bound}, that \eqref{eq_Perturbed-beta_strong} holds under the assumption
\begin{enumerate}
\renewcommand{\theenumi}{A\arabic{enumi}}
 \setcounter{enumi}{1}
	\item{
	\[0 < \min_{\substack{\bm x,\bm \beta: \|\bm x\|\leq  x_{\max},\\ \|\bm \beta\|\leq \beta_{\max}} } \dot{\mu}({ \bm \beta}^\top \bm x  )\,.\]
	}
	\label{ass_Perturbed-A2}
\end{enumerate}
Here $x_{\max}$ and $\beta_{\max}$ are the largest values of $\|\bm x\|$ and $\|{\bm \beta}\|$ for $t\geq 1$.

The above assumption holds for linear regressions and also for any model where the parameters $\hat{\bm{\beta}}_t$ remain bounded. We note that boundedness can be enforced through projection, for instance \cite{Filippi2010} took this approach. For the convergence rate \eqref{eq_Perturbed-beta_strong}, there are several alternative proofs on the rate of convergence of adaptive GLMs designs. These are discussed in the literature review. Here, any convergence result of the form \eqref{eq_Perturbed-beta_strong} can be used in place of Assumption \ref{ass_Perturbed-A2} and Proposition \ref{proposition_Perturbed-Beta-bound}. We provide our own proof under Assumption \ref{ass_Perturbed-A2} in order to present a short self-contained treatment.

\medskip

\noindent \emph{Time and Regret.}
Our goal to optimize revenue, \eqref{eq_Perturbed-maxR}, remains.
However, regrettably, we will always fall short of this goal. This is because the parameters $\bm{\beta}_{0}$ are unknown to us. 
Instead, we must simultaneously estimate $\bm{\beta}_{0}$ and choose prices that converge on $\bm{p}^{\star}(\c)$ for each $\bm{c}\in \mathcal{C}$. 
The variability in $\bm{x} = (\bm{p}, \c)$ required to estimate $\bm{\beta}_{0}$ will inevitably be detrimental to convergence towards $\bm{p}^{\star}(\c)$, while, rapid convergence in prices may inhibit estimation and lead to convergence to sub-optimal prices.
Stated more generally, there is a trade-off between exploration and exploitation which is well-known for bandit problems.

We let $T \in \mathbb{N}$ be the time horizon of our model.
For each time $t=1, \ldots, T$, we receive a vector of context $\bm{c}_{t} \in \mathcal{C}$. Then for $\bm{x}_{t}=\left(\bm{p}_{t}, \bm{c}_{t} \right)$, we are given response
\begin{equation}\label{eq_Perturbed-y_regression}
	y_{t} = \mu\left(\bm{\beta}_{0}^{\top}\bm{x}_{t} \right) + \varepsilon_{t}  \,,
\end{equation}
and receive reward $r_{t}\left(\bm{x}_{t}, \bm{y}_{t}\,; \bm{\beta}_0 \right)$ where
\[
r(\bm{x}_t\,;\bm{\beta}_0) 
=
\mathbb E [ r_t(\bm{x}_t, y_t \,; \bm{\beta}_0)\mid\bm{x}_t] \,.
\]
We let $\mathcal{F} = \{\mathcal{F}_t : t\in \mathbb Z_+\}$ be the filtration where $\mathcal{F}_t$ is the $\sigma$-field generated by random variables $(\bm{x}_s : s =1,\ldots,t)$.
Notice that $\bm{x}_t$ is $\mathcal{F}_{t-1}$ measurable.
We assume that $\epsilon_t$ defined in \eqref{eq_Perturbed-y_regression} is a martingale difference sequence w.r.t.\@ the filtration $\mathcal{F}_{t-1}$. 
That is $\mathbb E \left[ \epsilon_t \mid \mathcal{F}_{t-1} \right] = 0$ and for some constant $\sigma >0$ and $\gamma>2$, almost surely,
\begin{equation*}\label{eq_Perturbed-epsilon}
	\sup_t \mathbb E [ \epsilon_t^2 \mid \mathcal{F}_{t-1} ] = \sigma^2 < \infty 
	\,, \quad
	\sup_t \mathbb E [\mid\epsilon_t|^\gamma \mid \mathcal{F}_{t-1} ] < \infty \,. 
\end{equation*}
The \emph{regret} at time $T$, which we denote by $\mathcal{R}\! g(T)$, is defined by
\[
\mathcal{R}\! g(T)
:=
\sum_{t=1}^T 
r_t(\bm{x}^\star_t,y^\star_t \,; \bm{\beta}_0) - r_t(\bm{x}_t, y_t \,; \bm{\beta}_0)\,. 
\]
where $\bm{x}^{\star}_t := \left(\bm{p}^{\star}(\bm{c}_{t}), \bm{c}_{t}\right)$ and $y^\star_t$ is the response under given $\bm x^\star$.
 Recall that $\bm{p}^{\star}(\bm{c}_{t})$ is the optimal price for context $\bm{c}_{t}$  given parameter $\bm{\beta}_{0}$ is known, see \eqref{eq_Perturbed-maxR}.
The regret is the expected revenue loss from applying prices $\bm{p}_{t}$ rather than the optimal price $\bm{p}^{\star}$ when $\bm{\beta}_{0}$ is known. 
Thus as $\bm{\beta}_{0}$ is unknown, we instead look to make the regret as small as possible. 
As we discussed in the literature review, the best possible bounds on the regret for this class of problem are of the order $\O(\sqrt{T})$ (see \cite{kleinberg2003value}), and any policy that achieves regret $o(T)$  can be considered to have learned the optimal revenue. 

\subsection{Perturbed Certainty Equivalent Pricing}
The \emph{certainty equivalent price} is the price that treats all estimated parameters as if they were the true parameters of interest, and then chooses the optimal price given those estimates.
Specifically, for parameter estimates $\hat{\bm{\beta}}\in \R^d$, we define the certainty equivalent price $\bm{p}_{ce}(\bm{c}\,; \hat{\bm{\beta}})$ to be
\[
{\bm{p}}_{ce}(\bm{c}\,; \hat{\bm{\beta}})
    \in 
    \mathop{\arg\! \max}_{ \bm{p} \in \mathcal{P} } \;\; r(\bm{p}, \c\,; \hat{\bm{\beta}})\,.
\]
Notice this is exactly our optimization objective \eqref{eq_Perturbed-maxR}, with estimate $\hat{\bm \beta}$ in place of the true parameter $\bm{\beta}_0$.

For some control problems, the certainty equivalent rule can
be optimal, e.g.\@ linear quadratic control, and the certainty equivalent rule is a widely used scheme in such as model predictive control.
However, in general, it will lead to inconsistent learning and thus sub-optimal rewards and revenue \citep{LaiRobbins1982}.
Nonetheless, many companies will opt to use a certainty equivalent rule as it cleanly separates the problem of model learning from revenue optimization. 

With this in mind, we propose a simple implementable variant that maintains this separation. For parameter estimates $\bm{\beta}$, we choose prices 
\begin{equation*}
	\bm{p}  = \bm{p}_{ce} (\bm{c}\,; {\bm{\beta}}) + \alpha \bm{u} \,,
\end{equation*}
where $\bm u$ is an independent, bounded, mean zero random variable in $\mathbb R^m$ and $\alpha$ is a positive real number.
We call this the \emph{perturbed certainty equivalent price}.
Here, we simply add i.i.d.\@ noise to the certainty equivalent price in order to encourage exploration. 
Over time, we maintain the update rule
\begin{equation}\label{eq_Perturbed-perturb_p}
	\bm{p}_{t} = \bm{p}_{ce} ( \bm{c}_t \,; \hat{\bm{\beta}}_{t-1}) + \alpha_t \bm{u}_t \,,
\end{equation}
where $\bm{u}_{t}$ are i.i.d.\@ bounded, mean zero random vectors, $\hat{\bm{\beta}}_{t-1}$ is our current maximum (quasi-)likelihood estimate \eqref{eq_Perturbed-MQLE_beta}. 
Moreover, $\alpha_{t}$ is a deterministic decreasing sequence.
Shortly we will argue that taking $\alpha_t= t^{-\frac{1}{4}}$ is a good choice for achieving statistical consistency while achieving a good regret bounds.

Pseudo-code for the perturbed pricing algorithm is given in Algorithm~\ref{algo_Perturbed-PPA}, below. Here we split the procedure into 4 steps: \emph{context}, where we receive the context of the query and items to be sold; \emph{price}, where we select the perturbed certainty equivalent price; \emph{response}, where we receive information about the items sold and their revenue; and, \emph{estimate}, where we update our MQLE parameter. 
As discussed in the introduction, a key advantage of this scheme is its simplicity. 
Conventional algorithms involve deterministic penalties or confidence ellipsoids for choices close to the optimum. This in turn requires additional calculations such as matrix inversions and eigenvalue decomposition which modify the task of maximizing revenue and finding maximum likelihood estimators in a potentially non-trivial way. The proposed approach is appealing that the proposed algorithm maintains the statistical maximum likelihood objective and the revenue objective and the randomization added is a minor adjustment to the certainty equivalent rule.   

\begin{algorithm*}[H]
\textbf{Initialize:} $\eta>0$ and  
 $\hat{\bm \beta}_0 \in \mathcal B$  \\
\For{t=1,...,T}{
\underline{\textbf{Context}}: \\
Receive context $\bm{c}_t$.\\
\textbf{\underline{Price}}:\\
Choose price 
\begin{align*}
	\bm{p}_{t} &= \bm{p}_{ce} ( \bm{c}_t \,; \hat{\bm{\beta}}_{t-1}) + \alpha_t \bm{u}_t \,, &&
\end{align*}
where ${\bm{p}}_{ce}(\bm{c}\,; \hat{\bm{\beta}})\in  \mathop{\arg\! \max}_{ \bm{p} \in \mathcal{P} } r(\bm{p}, \c\,; \hat{\bm{\beta}})$, 
$\alpha_t = {t^{-\eta}}$, $\bm{u}_t$ is i.i.d.\@ mean zero \& covariance  $\Sigma_u \succ 0$.\\
\underline{\textbf{Response}}:\\
For input $\bm{x}_t = (\bm{p}_t, \bm{c}_t)$,\\
$\;\,$  receive response $y_t$,\\
$\;\,$ receive reward $r_t(\bm{x}_t, y_t\,; \bm{\beta}_0 )$.\\
\underline{\textbf{Estimate}}:\\
Calculate the MQLE $\hat{\bm \beta}_t$:
\begin{align*}
    &\sum_{s=1}^{t}
    \bm{x}_{s}
    \left(y_{s}-{\mu\left(\hat{\bm{\beta}}_{t}^{\top}\bm{x}_{s}\right)} \right)
    =\bm{0} \, . 	
&&
\end{align*}
}
 \caption{\textbf{Perturbed Pricing}\label{algo_Perturbed-PPA}} 
\end{algorithm*}

\section{Main Result}\label{section_Perturbed-result}
In this section, we present our main result, an upper bound on the regret under our policy.
Its proof is provided in Appendix A.

\begin{theorem}\label{theorem_Perturbed-regret-bound}
If $\alpha_t = t^{-\eta}$ for $\eta \in [1/4,1/2)$ then, with probability $1$,
the regret over time horizon $T$ is
\begin{align*}
    \begin{split}
        \mathcal{R} \! g(T)
        =   \mathcal{O} \left(\sqrt{T\log T}+T^{1-2\eta} +\sum_{t=1}^{T}\left(\frac{\log(t)}{t^{1-2\eta}}\right)\right) \,.
    \end{split}
\end{align*}
Choosing $\eta = \frac{1}{4}$ gives
\begin{align*}
    \begin{split}
        \mathcal{R} \! g(T)
        & =   \mathcal{O} \left(\sqrt{T} \log (T)\right)\,.
    \end{split}
\end{align*}
\end{theorem}
The order of the regret bound above is consistent with prior results such as \citet{DenBoerZwart2014S} and \citet{KeskinZeevi2014} which achieve a bound of the same order.

We now describe the results that are required to prove Theorem \ref{theorem_Perturbed-regret-bound}, along with some notation.

\subsection{Additional Notation}\label{sec_additionalnotation}
For a vector $\bm{x} \in \mathbb{R}^{d}$, we let $\|\bm{x}\|$ denote its Euclidean norm and $\|\bm{x}\|_\infty$ denote its supremum norm.
Because we wish to consider the design matrix $\sum_{s=1}^{t} \bm{x}_{s} \bm{x}_{s}^\top$, we add some notation needed to re-express the effect of perturbation on the price. 
Specifically, we re-express the perturbed certainty equivalent rule (price) in terms of the full input vector $\bm{x}= (\bm{p}, \bm{c})$ rather than just in terms of $\bm{p}$. 
Specifically, we let 
\[
\bm{x}_{t} = \hat{\bm{x}}_{t} + \alpha_t \bm{z}_{t}\,,
\]
where
\[
\hat{\bm{x}}_{t} = \left(\bm{p}_{ce}\left(\bm{c}_t \,; \hat{\bm{\beta}}_{t}\right), \bm{c}_t \right)\,,
\]
and 
\[
\bm{z}_t = \left(\bm{u}_t, \bm{0} \right) \in \mathbb{R}^{d}\,.
\]
Given our boundedness assumptions, we apply the notation $p_{\max}, c_{\max} \in \mathbb{R}^{+}$ where $\left\|\bm{p}_t\right\|_{\infty} \leq p_{\max}$ and $\left\| \bm{c}_t\right\|_{\infty} \leq c_{\max}$ for all $t \in  \mathbb{Z}_{+}$.
Recall that $\bm{u}_{t}$ are vectors of i.i.d.\@ bounded random variables, and therefore so are $\bm{z}_{t}$.
We assume that $\left\|\bm{u}_t\right\|_{\infty} \leq u_{\max}$ for all $t \in  \mathbb{Z}_{+}$ where $u_{\max}\in \mathbb{R}_{+}$, thus there exists $z_{\max}\in \mathbb{R}_{+}$ such that  $\left\|\bm z_t\right\|_{\infty} \leq {z}_{\max}$ for all $t \in  \mathbb{Z}_{+}$.
We denote the covariance matrix of $\bm{z}_{t}$ by $\Sigma^{z} = \mathbb{E}\left[\bm{z}_{t} \bm{z}_{t}^{\top}\right]$.
Furthermore, we let $\lambda_{\max} (t)$ and $\lambda_{\min} (t)$ denote the maximum and minimum eigenvalues of $\sum_{s=1}^{t}\bm{{x}}_s \bm{{x}}_s ^{\top}$.

\subsection{Key Additional Results}
To prove Theorem \ref{theorem_Perturbed-regret-bound}, we require additional results: Lemma \ref{lemma_Perturbed-order_r}, Proposition \ref{proposition_Perturbed-Beta-bound} and Proposition \ref{proposition_Perturbed-lambdamin-bound}, which are stated below, and their proofs are given in Appendix B.

Lemma \ref{lemma_Perturbed-order_r} shows that the performance of the perturbed policy depends on how it learns the unknown parameter $\bm{\beta}$.
\begin{lemma}\label{lemma_Perturbed-order_r}
	For $\bm{p}^{\star} \in \mathop{\arg\! \max}_{\bm{p}\in \mathcal{P}} \; r\left(\bm{p}, \bm{c}\,; \bm{\beta}_{0}\right)$, there exists $K_{0}>0$  such that, for all $\bm{p} \in \mathcal{P}$ and $\bm{c}\in \mathcal{C}$
	\begin{equation}\label{eq_Perturbed-r_bound_p}
    	\left|r\left(\bm{p}, \bm{c}\,; \bm{\beta}_0\right)-r\left(\p^{\star}, \bm{c}\,; \bm{\beta}_{0}\right)\right|
    	\leq K_{0} \left\|\bm{p}-\bm{p}^{\star}\right\|^2 \, .
	\end{equation}
If either Assumption~\ref{ass_Perturbed-A_finite} or \ref{ass_Perturbed-A_convex} holds, then there exists $K_{1}>0$ such that 
    \begin{equation}\label{eq_Perturbed-p_bound_beta}
    	\sup_{\bm{c}\in\mathcal{C}}\left\|\bm{p}^{\star}(\bm{c}\,; \bm{\beta})-\bm{p}^{\star}(\bm{c})\right\|
    	\leq K_{1}
    	\left\|{\bm{\beta}} - \bm{\beta}_{0}\right\|\,.
    \end{equation}
\end{lemma}
This lemma establishes the continuity of revenue as a function of parameters $\bm{\beta}$.
The regret is shown to be the same order of the squared feature vectors $\|\p-\p^{*}\|^2$, which is equal to that of the squared error of the estimated parameters $\| \hat{\bm{\beta}}_t - \bm{\beta}_{0}\|^2$. 
This result applies a combination of the Lipschitz continuity of $r(\bm{x} \,; \bm{\beta}_0)$ and the Implicit Function Theorem (Theorem 9.28 of \cite{rudin1976principles}). 

Proposition \ref{proposition_Perturbed-Beta-bound} gives the strong consistency result for GLMs.
\begin{proposition}\label{proposition_Perturbed-Beta-bound}
Under Assumption \ref{ass_Perturbed-A2}, as $t \to \infty$ with probability $1$,
\[
\left\| \hat{\bm{\beta}}_{t} - {\bm{\beta}}_{0}\right \|^2 
= \O  \left(\frac{\log(t)}{\lambda_{\min}(t)}\right) \,.
\]
\end{proposition}
Along with Lemma \ref{lemma_Perturbed-order_r}, Proposition \ref{proposition_Perturbed-Beta-bound} establishes that the key quantity to determining the regret is $\lambda_{\min}(t)$.
As discussed in Section \ref{section_Perturbed-review}, under Assumption \ref{ass_Perturbed-A2}, there are a number of similar results which can be used in place of Proposition \ref{proposition_Perturbed-Beta-bound}. The argument presented initially takes the approach of \cite{ChenHY1999} and \cite{Li2017} and then applies Lemma 1iii) of \cite{LaiWei1982}.

We now construct a lower bound for the smallest eigenvalue $\lambda_{\min}(t)$ as follows.

\begin{proposition}\label{proposition_Perturbed-lambdamin-bound}
If $\alpha_s = s^{-\eta}$ where $s = 1, \ldots, t$, for $\eta \in [0,1/2)$ then
		\begin{align*}
			\lambda_{\min} \left(t\right)
			= \Omega \left(\sum_{s=1}^{t}\alpha_{s}^{2}\right) \,.
		\end{align*}
\end{proposition}
The above proposition applies a new bound on minimum eigenvalue of the design matrix $\sum_{s=1}^{t}\bm{{x}}_s \bm{{x}}_s ^{\top}$, which is critical to our proof.
This algebraic eigenvalue bound is given in Proposition \ref{proposition_Perturbed-lambdaM}, and proof of which is in Appendix C.
This new eigenvalue bound is obtained based on decomposition according to the Schur Complement. A related eigenvalue bounds is Ostrowski's Theorem, see \cite{horn2012matrix}. 

\section{Numerical Experiments}\label{section_Perturbed-experiments}
In this section, we examine the performance of the perturbed certainty equivalent pricing developed in the cases of linear and logistic demand models. This short numerical study confirms the behavior proved above. We consider moderate sized problems with 17 explanatory variables. We note that problems of this size can be achieved from large practical problems through dimension reduction. We do not investigate such reductions here, instead we refer the reader to the seminal work \cite{Li2010} where a recommendation system with several thousand features is reduced to a problem of 36 explanatory variables and a Linear UCB model first applied. 

In our simulation experiments, we consider a company that sells a single product and would like to decide pricing policy based on previous selling prices and $15$ other contextual variables. 
At each time $t$, we select a selling price $p_t \in [{p}_{l}, {p}_{h}]$, and receive a context vector $\c_{t}$. 
We set the feasible price $[{p}_{l}, {p}_{h}] = [0.5, 5]$. 
The context vector $\c_{t}$ are multivariate normal random variables with $m = 15$ dimensions, means equal to a vector of zeros and covariance matrix $\Sigma^{c} = I_{15}$, where $I_{15}$ is a $15 \times 15$ identity matrix.
Then, we obtain the feature vector ${\bm{x}}_{t} = \left(\bm{p}_{t}, \c_t \right) \in \mathbb{R}^{17}$, where $\bm{p}_{t} = (1, p_t)$.
We assume that the true values of unknown parameters of price vector are $(\beta_0, \beta_1) =(1, -0.5)^{\top} \in \mathbb{R}^{2}$, and $\c_t$ and true coefficients $\beta_2,...,\beta_{16}$ are drawn i.i.d.\@ from a multivariate Gaussian distribution $N(0,I)\in \mathbb{R}^{15}$.
We model the revenue as the price times the demand responses, i.e. $r_t(\bm{x}_t, y_t)=p_t y_t$ and $r(\hat \x_t;\bbeta)= \mathbb E [ r_t(\bm{x}_t, y_t)\mid \bm{x}_t]$.
We simulate our policy for $T=2000$ with $\alpha^{t} = t^{-1/4}$ and measure the performance of the policy by regret.

\begin{figure}[t]
     \centering
     \begin{subfigure}[b]{0.49\textwidth}
         \centering
         \includegraphics[width=\textwidth]{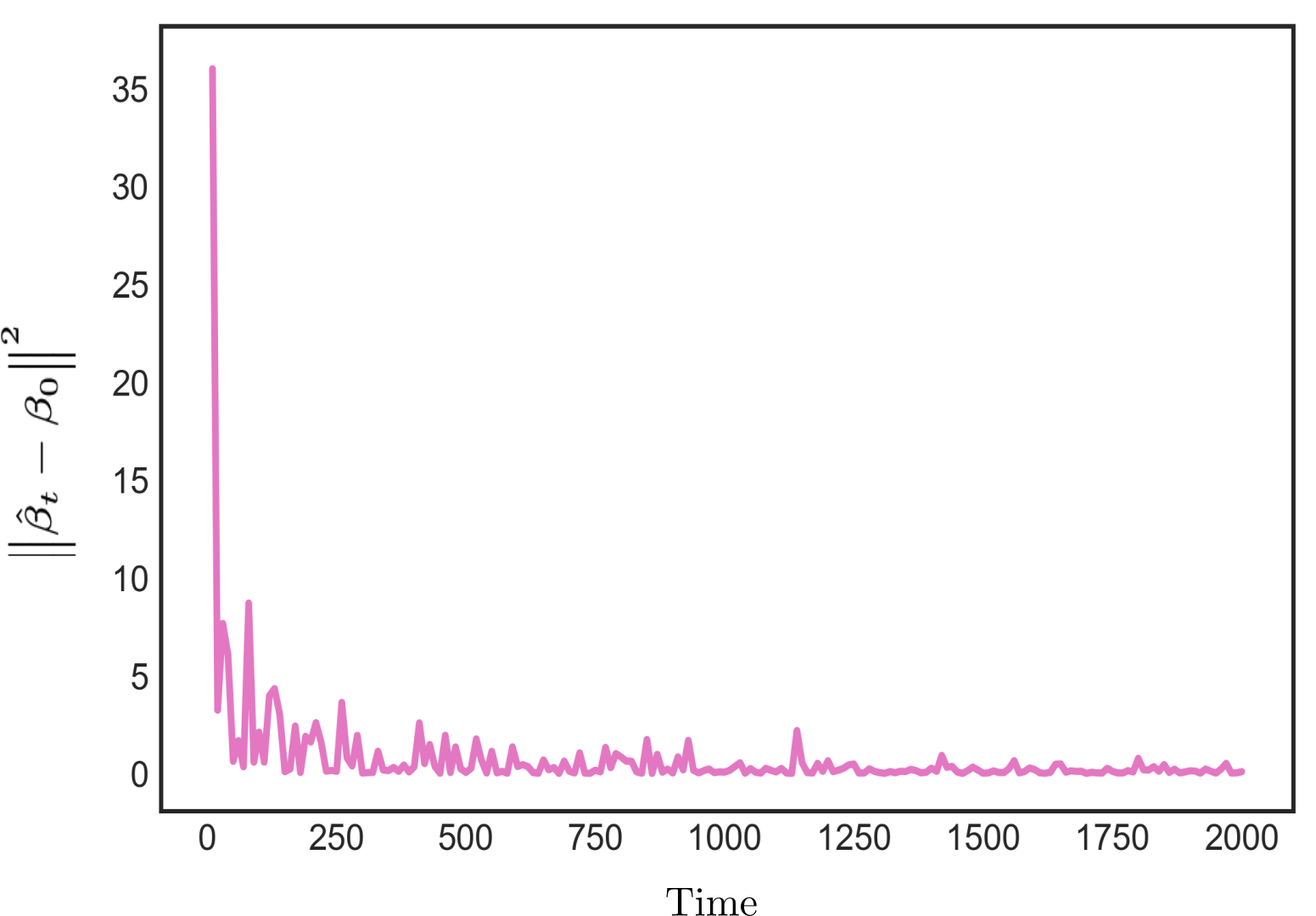}
         \caption{$\|\hat{\bm{\beta}}_{t} - {\bm{\beta}}_{0} \|^2$}
         \label{fig:Perturbed_linear_beta_bound}
     \end{subfigure}
     \hfill
     \begin{subfigure}[b]{0.49\textwidth}
         \centering
         \includegraphics[width=\textwidth]{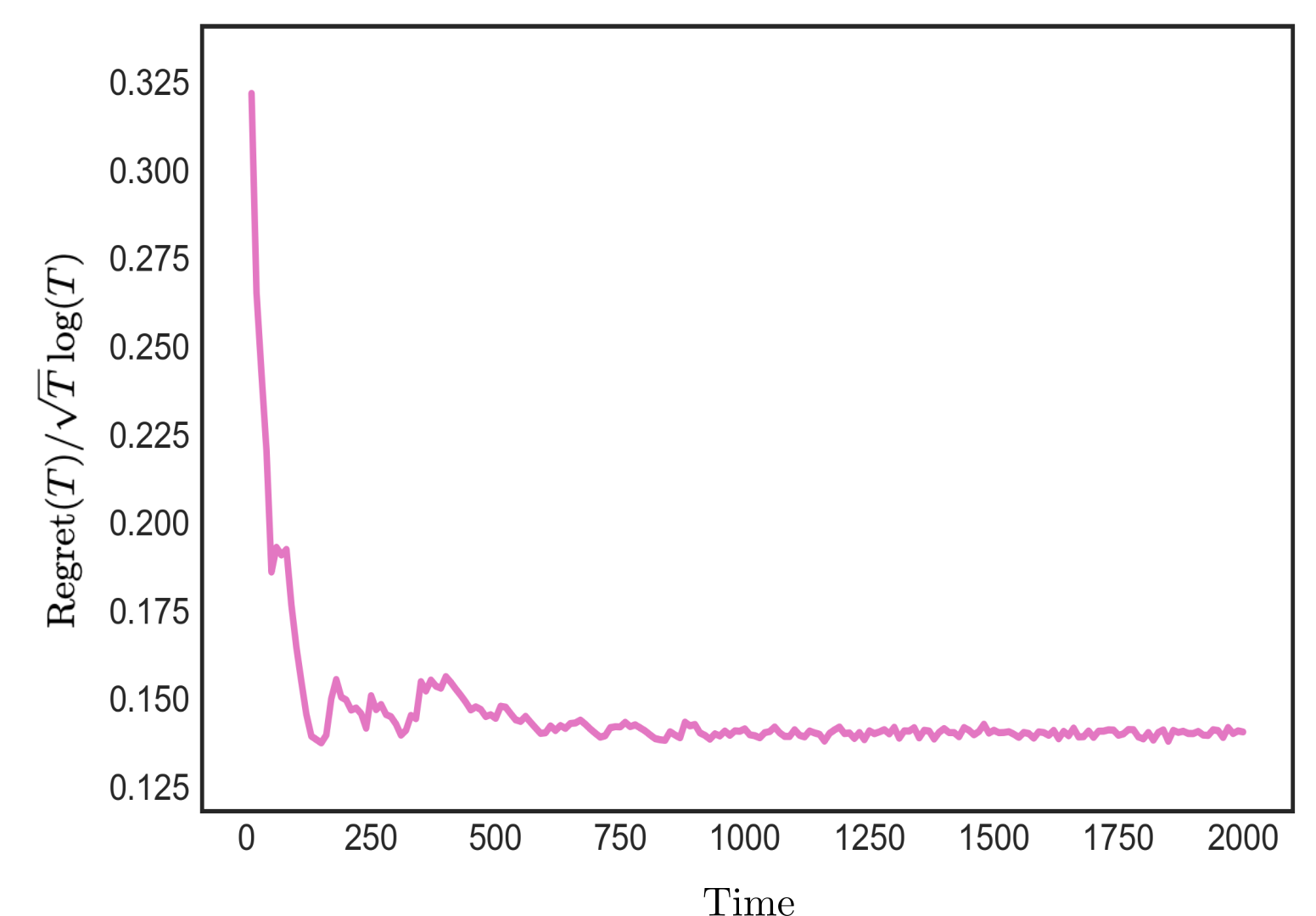}
         \caption{$\mathcal R\! g(T)/\sqrt{T}\log(T)$}
         \label{fig:Perturbed_linear_regret_ratio}
     \end{subfigure}
    \caption{Convergence of parameter estimates and regret with linear demand function. Time period $T=2000$ and $\alpha^{t} = t^{-1/4}$. Problem parameters used are $[{p}_{l}, {p}_{h}] = [0.5, 2]$, $m = 15$, true parameters of price vector $[1, -0.5]^{\top} \in \mathbb{R}^{2}$,  $\c_t$ and its true coefficients drawn i.i.d.\@ from $N(0,I)\in \mathbb{R}^{15}$.}
\end{figure}

\begin{figure}
     \centering
     \begin{subfigure}[b]{0.49\textwidth}
         \centering
         \includegraphics[width=\textwidth]{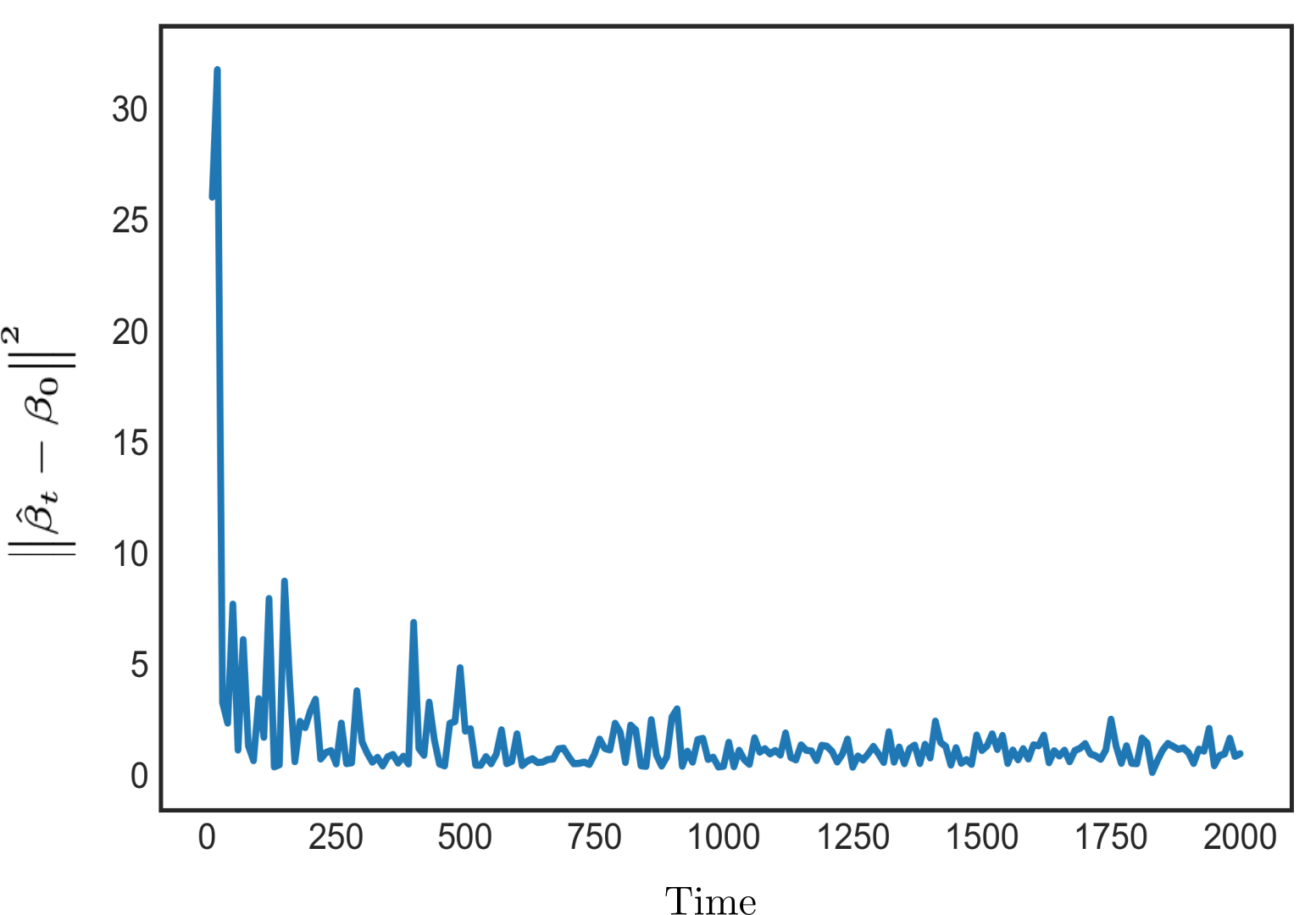}
         \caption{$\| \hat{\bm{\beta}}_{t} - {\bm{\beta}}_{0} \|^2$}
         \label{fig:Perturbed_logistic_beta_bound}
     \end{subfigure}
     \hfill
     \begin{subfigure}[b]{0.49\textwidth}
         \centering
         \includegraphics[width=\textwidth]{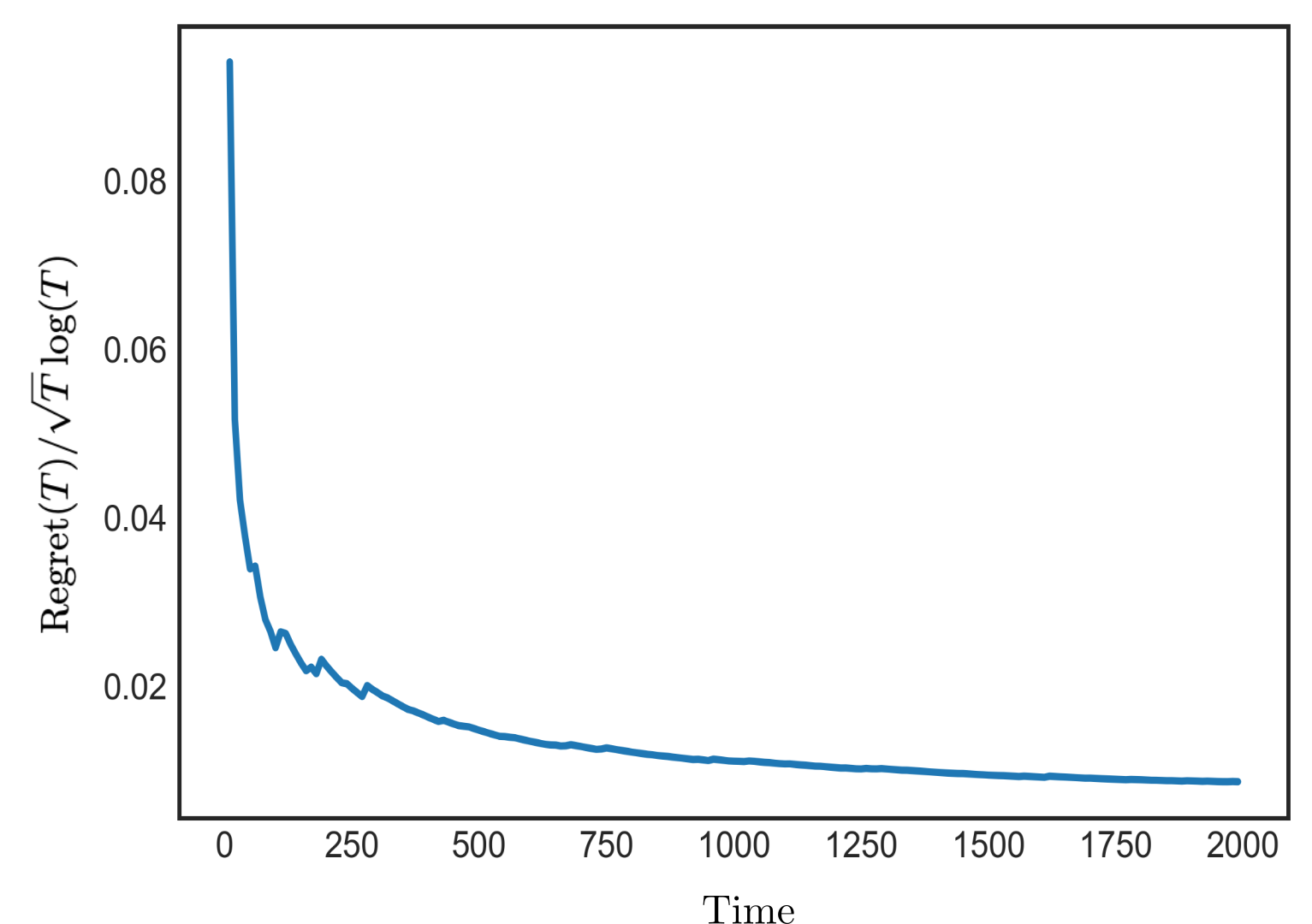}
         \caption{$\mathcal R\! g(T)/\sqrt{T}\log(T)$}
         \label{fig:Perturbed_logistic_regret_ratio}
     \end{subfigure}
    \caption{Convergence of parameter estimates and regret with logistic regression for demand. Time period $T=2000$ and $\alpha^{t} = t^{-1/4}$. Problem parameters used are $[{p}_{l}, {p}_{h}] = [0.5, 2]$, $m = 15$, true parameters of price vector $[1, -0.5]^{\top} \in \mathbb{R}^{2}$,  $\c_t$ and its true coefficients drawn i.i.d.\@ from $N(0,I)\in \mathbb{R}^{15}$.}
\end{figure}

We first consider that demand follows a linear function, i.e. a GLM with an identity link function $\mu(x) = x$.
Figure \ref{fig:Perturbed_linear_beta_bound} shows the value of $\| \hat{\bm{\beta}}_{t} - {\bm{\beta}}_{0} \|^2$, which is the squared norm of the difference between the parameter estimates $\hat{\bm{\beta}}_{t}$ and the true parameters $\bbeta_{0}$.
We can see that the difference converges to zero as $t$ becomes large, which demonstrates the strong consistency of parameter estimates. 
To show the order of regret, we re-scale it to $\mathcal R\! g(T)/\sqrt{T} \log(T)$ in Figure \ref{fig:Perturbed_linear_regret_ratio}.
It shows that the $\mathcal R\! g(T)/\sqrt{T}\log(T)$ converges to a constant value as $T$ becomes large, which verifies that the regret has, at most, an order of $\sqrt{T} \log(T)$ under our policy. The constant in this case is quite small close to $0.14$.
In practice, the response that seller receives may be a zero-one response corresponding to the item being unsold or sold. Here a logistic regression model is appropriate, with link function $\mu(x) = 1/(1+\e^{-x})$.
Figures~\ref{fig:Perturbed_logistic_beta_bound} shows that $\| \hat{\bm{\beta}}_{t} - {\bm{\beta}}_{0} \|^2$ converges to zero as $t$ becomes sufficiently large and Figure \ref{fig:Perturbed_logistic_regret_ratio} shows that regret is of order $\sqrt{T} \log(T)$.
Again the constant in the regret is small, around 0.01, suggesting good dependence on the size of this problem.
\section{Discussion, Conclusions and Future Directions}\label{section_Perturbed-discussion}
We considered a dynamic learning and contextual pricing problem with unknown demand.
This work acts to provide a theoretical foundation to positive empirical findings in \cite{lobo2003pricing} and finds the correct magnitude of perturbation for the best regret scaling.  We allow for contextual information in the pricing decision. 

We focus on a maximum likelihood estimation approach. However, for high dimensional problems, where dimension reduction is not possible,
a stochastic gradient descent approach must instead be used for parameter estimation. Thus one direction of future research is to consider the setting where parameter estimates $\hat{\bm{\beta}}_{t}$ and prices $p_t$ are updated online according to a Robbins--Monro rule. We focus on i.i.d. contexts, but it should be clear from the proofs that we may allow contexts to evolve in a more general adaptive manner, so long variability in contextual information dominates variability in prices.
Also, we focus on i.i.d.\@ perturbation, but other forms of perturbation could be considered. For instance, Quasi-Monte-Carlo methods can reduce variance while more systematically exploring the space perturbations. We allow the size of perturbations to decrease uniformly over each context. However, in practice one might want to let this decrease depending on the number of times that a context has occurred. Thus the asynchronous implementation of the approach is an important consideration.
We consider a single retailer selling multiple items over multiple contexts. However, adversarial competition is often important. For instance the interplay between large sets of contexts and a variety of sellers occurs in online advertising auctions. Here an understanding of regret in both the stochastic and adversarial setting would be an important research direction.

Finally to summarize, certainty equivalent pricing is commonly applied in management science. Random perturbation around the certainty equivalent price is simple and when combined with standard statistical parameter estimation it achieves revenue comparable with more technically sophisticated learning algorithms. 

%
\subsection*{ACKNOWLEDGMENT}{Yuqing Zhang gratefully acknowledges the support of the China Scholarship Council. Neil Walton thanks Nick Higham for discussions and pointers that lead to the eigenvalue decomposition in Proposition 3, and Arnoud den Boer for commenting on an early draft of this manuscript.}

\bibliographystyle{abbrvnat}
\bibliography{mybibfile}

\begin{thebibliography}{43}
\providecommand{\natexlab}[1]{#1}
\providecommand{\url}[1]{\texttt{#1}}
\expandafter\ifx\csname urlstyle\endcsname\relax
  \providecommand{\doi}[1]{doi: #1}\else
  \providecommand{\doi}{doi: \begingroup \urlstyle{rm}\Url}\fi

\bibitem[Abbasi-Yadkori et~al.(2011)Abbasi-Yadkori, P\'{a}l, and
  Szepesv\'{a}ri]{AbbasiYadkori2011}
Y.~Abbasi-Yadkori, D.~P\'{a}l, and C.~Szepesv\'{a}ri.
\newblock Improved algorithms for linear stochastic bandits.
\newblock In \emph{Proceedings of the 24th International Conference on Neural
  Information Processing Systems (NIPS)}, pages 2312--2320, 2011.
\newblock URL
  \url{http://david.palenica.com/papers/linear-bandit/linear-bandits-NIPS2011-camera-ready.pdf}.

\bibitem[Amin et~al.(2014)Amin, Rostamizadeh, and Syed]{Amin2014}
K.~Amin, A.~Rostamizadeh, and U.~Syed.
\newblock Repeated contextual auctions with strategic buyers.
\newblock In \emph{Advances in Neural Information Processing Systems 27 (NIPS
  2014)}, 2014.
\newblock URL
  \url{http://amin.kareemx.com/pubs/AminRostamizadehSyedNIPS2014.pdf}.

\bibitem[Anderson and Taylor(1976)]{AndersonTaylor1976}
T.~W. Anderson and J.~B. Taylor.
\newblock Strong consistency of least squares estimates in normal linear
  regression.
\newblock \emph{The Annals of Statistics}, 4\penalty0 (4):\penalty0 788--790,
  1976.
\newblock \doi{10.1214/aos/1176343552}.

\bibitem[Auer(2003)]{Auer2002a}
P.~Auer.
\newblock Using confidence bounds for exploitation-exploration trade-offs.
\newblock \emph{Journal of Machine Learning Research}, 3\penalty0 (3):\penalty0
  397--422, 2003.
\newblock \doi{10.1109/SFCS.2000.892116}.

\bibitem[Ban and Keskin(2020)]{BanKeskin2019}
G.-Y. Ban and N.~B. Keskin.
\newblock Personalized dynamic pricing with machine learning: High dimensional
  features and heterogeneous elasticity.
\newblock \emph{Management Science}, 2020.
\newblock forthcoming.

\bibitem[Besbes and Zeevi(2009)]{BesbesZeevi2009}
O.~Besbes and A.~Zeevi.
\newblock Dynamic pricing without knowing the demand function: Risk bounds and
  near-optimal algorithms.
\newblock \emph{Operations Research}, 57\penalty0 (6):\penalty0 1407--1420,
  2009.
\newblock URL \url{https://doi.org/10.1287/opre.1080.0640}.

\bibitem[Broder and Rusmevichientong(2012)]{BroderRusmevichientong2012}
J.~Broder and P.~Rusmevichientong.
\newblock Dynamic pricing under a general parametric choice model.
\newblock \emph{Operations Research}, 60\penalty0 (4):\penalty0 965--980, July
  2012.
\newblock URL \url{http://search.proquest.com/docview/1041256005/}.

\bibitem[Bubeck and Cesa{-}Bianchi(2012)]{BubeckBianchi2012}
S.~Bubeck and N.~Cesa{-}Bianchi.
\newblock Regret analysis of stochastic and nonstochastic multi-armed bandit
  problems.
\newblock \emph{CoRR}, 5\penalty0 (1):\penalty0 1--122, 2012.
\newblock \doi{10.1561/2200000024}.

\bibitem[Chang(1999)]{Chang1999}
I.~Chang, Y-c.
\newblock Strong consistency of maximum quasi-likelihood estimate in
  generalized linear models via a last time.
\newblock \emph{Statistics \& Probability Letters}, 45\penalty0 (3):\penalty0
  237--246, 1999.

\bibitem[Chen et~al.(1999)Chen, Hu, and Ying]{ChenHY1999}
K.~Chen, I.~Hu, and Z.~Ying.
\newblock Strong consistency of maximum quasi-likelihood estimators in
  generalized linear models with fixed and adaptive designs.
\newblock \emph{The Annals of Statistics}, 27\penalty0 (4):\penalty0
  1155--1163, 1999.
\newblock \doi{10.1214/aos/1017938919}.

\bibitem[Chen et~al.(2015)Chen, Owen, Pixton, and
  Simchi-Levi]{chen2015statistical}
X.~Chen, Z.~Owen, C.~Pixton, and D.~Simchi-Levi.
\newblock A statistical learning approach to personalization in revenue
  management.
\newblock \emph{Available at SSRN 2579462}, 2015.

\bibitem[Chu et~al.(2011)Chu, Li, Reyzin, and Schapire]{Chu2011}
W.~Chu, L.~Li, L.~Reyzin, and R.~Schapire.
\newblock Contextual bandits with linear payoff functions.
\newblock In \emph{Proceedings of the Fourteenth International Conference on
  Artificial Intelligence and Statistics}, volume~15, pages 208--214, 2011.
\newblock URL \url{http://proceedings.mlr.press/v15/chu11a.html}.

\bibitem[Cohen et~al.(2016)Cohen, Lobel, and Paes~Leme]{Cohen2016}
M.~Cohen, I.~Lobel, and R.~Paes~Leme.
\newblock Feature-based dynamic pricing.
\newblock \emph{ACM Conference on Economics \& Computation (EC)}, 15\penalty0
  (2):\penalty0 40--44, 2016.
\newblock URL \url{http://dx.doi.org/10.2139/ssrn.2737045}.

\bibitem[Dani et~al.(2008)Dani, Hayes, and Kakade]{Dani2008}
V.~Dani, T.~P. Hayes, and S.~M. Kakade.
\newblock Stochastic linear optimization under bandit feedback.
\newblock In \emph{21st Annual Conference on Learning Theory (COLT)}, pages
  355--366, 2008.
\newblock URL \url{http://colt2008.cs.helsinki.fi/papers/80-Dani.pdf}.

\bibitem[den Boer(2013)]{DenBoer2013}
A.~den Boer.
\newblock \emph{Dynamic Pricing and Learning}.
\newblock PhD thesis, VU University of Amsterdam, 2013.
\newblock URL
  \url{http://dare.ubvu.vu.nl/bitstream/handle/1871/39660/dissertation.pdf}.

\bibitem[den Boer(2015)]{DenBoer2015}
A.~den Boer.
\newblock Dynamic pricing and learning: Historical origins, current research,
  and new directions.
\newblock \emph{Surveys in operations research and management science},
  20\penalty0 (1):\penalty0 1--18, 2015.
\newblock \doi{10.1016/j.sorms.2015.03.001}.

\bibitem[den Boer and Zwart(2014{\natexlab{a}})]{DenBoerZwart2014S}
A.~den Boer and B.~Zwart.
\newblock Simultaneously learning and optimizing using controlled variance
  pricing.
\newblock \emph{Management Science}, 60\penalty0 (3):\penalty0 770--783,
  2014{\natexlab{a}}.
\newblock \doi{10.1287/mnsc.2013.1788}.

\bibitem[den Boer and Zwart(2014{\natexlab{b}})]{denBoerZwart2014M}
A.~den Boer and B.~Zwart.
\newblock Mean square convergence rates for maximum quasi- likelihood
  estimators.
\newblock \emph{Stochastic Systems}, 4\penalty0 (2):\penalty0 375--403,
  2014{\natexlab{b}}.
\newblock \doi{10.1287/12-SSY086}.

\bibitem[Filippi et~al.(2010)Filippi, Cappe, Garivier, and
  Szepesv\'{a}ri]{Filippi2010}
S.~Filippi, O.~Cappe, A.~Garivier, and C.~Szepesv\'{a}ri.
\newblock Parametric bandits: The generalized linear case.
\newblock In \emph{Advances in Neural Information Processing Systems 23 (NIPS
  2010)}, pages 586--594, 2010.
\newblock URL
  \url{https://sites.ualberta.ca/~szepesva/papers/GenLinBandits-NIPS2010.pdf}.

\bibitem[Gallego and Van~Ryzin(1994)]{gallego1994optimal}
G.~Gallego and G.~Van~Ryzin.
\newblock Optimal dynamic pricing of inventories with stochastic demand over
  finite horizons.
\newblock \emph{Management science}, 40\penalty0 (8):\penalty0 999--1020, 1994.

\bibitem[Horn and Johnson(2012)]{horn2012matrix}
R.~A. Horn and C.~R. Johnson.
\newblock \emph{Matrix analysis}.
\newblock Cambridge university press, 2012.

\bibitem[Javanmard and Nazerzadeh(2019)]{JavanmardNazerzadeh2019}
A.~Javanmard and H.~Nazerzadeh.
\newblock Dynamic pricing in high-dimensions.
\newblock \emph{Journal of Machine Learning Research}, 20\penalty0
  (9):\penalty0 1--49, 2019.
\newblock URL \url{http://jmlr.org/papers/v20/17-357.html}.

\bibitem[Keskin and Zeevi(2014)]{KeskinZeevi2014}
N.~B. Keskin and A.~Zeevi.
\newblock Dynamic pricing with an unknown demand model: Asymptotically optimal
  semi-myopic policies.
\newblock \emph{Operations Research}, 62\penalty0 (5):\penalty0 1142--1167,
  2014.
\newblock \doi{10.1287/opre.2014.1294}.

\bibitem[Kleinberg and Leighton(2003)]{kleinberg2003value}
R.~Kleinberg and T.~Leighton.
\newblock The value of knowing a demand curve: Bounds on regret for online
  posted-price auctions.
\newblock In \emph{44th Annual IEEE Symposium on Foundations of Computer
  Science, 2003. Proceedings.}, pages 594--605. IEEE, 2003.

\bibitem[Lai(2003)]{Lai2003}
T.~Lai.
\newblock Stochastic approximation: invited paper.
\newblock \emph{The Annals of Statistics}, 31\penalty0 (2):\penalty0 391--406,
  2003.
\newblock \doi{10.1214/aos/1051027873}.

\bibitem[Lai and Robbins(1982)]{LaiRobbins1982}
T.~Lai and H.~Robbins.
\newblock Iterated least squares in multiperiod control.
\newblock \emph{Advances in Applied Mathematics}, 3\penalty0 (1):\penalty0
  50--73, 1982.
\newblock \doi{10.1016/S0196-8858(82)80005-5}.

\bibitem[Lai and Wei(1982)]{LaiWei1982}
T.~L. Lai and C.~Z. Wei.
\newblock Least squares estimates in stochastic regression models with
  applications to identification and control of dynamic systems.
\newblock \emph{The Annals of Statistics}, 10\penalty0 (1):\penalty0 154--166,
  1982.
\newblock \doi{10.1214/aos/1176345697}.

\bibitem[Lattimore and Szepesv\'{a}ri(2017)]{LattimorSzepesvsrie2016}
T.~Lattimore and C.~Szepesv\'{a}ri.
\newblock The end of optimism? an asymptotic analysis of finite-armed linear
  bandits.
\newblock In \emph{Proceedings of the 20th International Conference on
  Artificial Intelligence and Statistics, {AISTATS} 2017, 20-22 April 2017,
  Fort Lauderdale, FL, {USA}}, pages 728--737, 2017.
\newblock URL \url{http://proceedings.mlr.press/v54/lattimore17a.html}.

\bibitem[Li et~al.(2010)Li, Chu, Langford, and Schapire]{Li2010}
L.~Li, W.~Chu, J.~Langford, and R.~E. Schapire.
\newblock A contextual-bandit approach to personalized news article
  recommendation.
\newblock In \emph{Proceedings of the 19th International Conference on World
  Wide Web}, WWW'10, page 661–670. Association for Computing Machinery, 2010.
\newblock \doi{10.1145/1772690.1772758}.

\bibitem[Li et~al.(2017)Li, Lu, and Zhou]{Li2017}
L.~Li, Y.~Lu, and D.~Zhou.
\newblock Provably optimal algorithms for generalized linear contextual
  bandits.
\newblock In \emph{Proceedings of the 34th International Conference on Machine
  Learning}, volume~70, pages 2071--2080, 2017.
\newblock URL \url{http://proceedings.mlr.press/v70/li17c.html}.

\bibitem[Lobo and Boyd(2003)]{lobo2003pricing}
M.~S. Lobo and S.~Boyd.
\newblock Pricing and learning with uncertain demand.
\newblock In \emph{INFORMS Revenue Management Conference}, 2003.

\bibitem[McCullagh and Nelder(1989)]{McCullaghNelder1989}
P.~McCullagh and J.~A. Nelder.
\newblock \emph{Generalized Linear Models}.
\newblock Chapman and Hall, London, 2nd edition, 1989.
\newblock ISBN 978-0412317606.

\bibitem[Ohlsson and Johansson(2010)]{OhlssonJohansen2010}
E.~Ohlsson and B.~Johansson.
\newblock \emph{Non-Life Insurance Pricing with Generalized Linear Models}.
\newblock Springer, Berlin, Heidelberg, 2010.

\bibitem[Phillips(2005)]{PhillipsRobert2005}
R.~L. Phillips.
\newblock \emph{Pricing and revenue optimization}.
\newblock Stanford Business Books, Stanford, Calif., 2005.
\newblock ISBN 0804746982.

\bibitem[Qiang and Bayati(2016)]{Qiang2016}
S.~Qiang and M.~Bayati.
\newblock Dynamic pricing with demand covariates.
\newblock Working paper, Stanford University Graduate School of Business,
  Stanford, CA, 2016.
\newblock URL \url{http://web.stanford.edu/~bayati/papers/dpdc.pdf}.

\bibitem[Rudin(1976)]{rudin1976principles}
W.~Rudin.
\newblock \emph{Principles of Mathematical Analysis}.
\newblock International series in pure and applied mathematics. McGraw-Hill,
  1976.
\newblock ISBN 9780070856134.
\newblock URL \url{https://books.google.co.uk/books?id=kwqzPAAACAAJ}.

\bibitem[Rusmevichientong and Tsitsiklis(2010)]{RusmevichientongTsitsiklis2010}
P.~Rusmevichientong and J.~N. Tsitsiklis.
\newblock Linearly parameterized bandits.
\newblock \emph{Mathematics of Operations Research}, 35\penalty0 (2):\penalty0
  395--411, 2010.
\newblock \doi{10.1287/moor.1100.0446}.

\bibitem[Talluri and van Ryzin(2005)]{TalluriRyzin2005}
K.~Talluri and G.~van Ryzin.
\newblock \emph{The theory and practice of revenue management}.
\newblock Springer, Boston, MA, 2005.
\newblock ISBN 1-280-46175-6.
\newblock \doi{10.1007/b139000}.

\bibitem[Vershynin(2012)]{vershynin2012close}
R.~Vershynin.
\newblock How close is the sample covariance matrix to the actual covariance
  matrix?
\newblock \emph{Journal of Theoretical Probability}, 25\penalty0 (3):\penalty0
  655--686, 2012.

\bibitem[Vershynin(2018)]{Vershynin2018}
R.~Vershynin.
\newblock \emph{High-Dimensional Probability: An Introduction with Applications
  in Data Science}.
\newblock Cambridge Series in Statistical and Probabilistic Mathematics.
  Cambridge University Press, 2018.
\newblock ISBN 9781108415194.
\newblock URL
  \url{https://www.math.uci.edu/~rvershyn/papers/HDP-book/HDP-book.pdf}.

\bibitem[Wedderburn(1974)]{Wedderburn1974}
R.~W.~M. Wedderburn.
\newblock Quasi-likelihood functions, generalized linear models, and the
  gauss-newton method.
\newblock \emph{Biometrika}, 61\penalty0 (3):\penalty0 439--447, 1974.
\newblock \doi{10.2307/2334725}.

\bibitem[Williams(1991)]{Williams1991}
D.~Williams.
\newblock \emph{Probability with Martingales}.
\newblock Cambridge mathematical textbooks. Cambridge University Press, 1991.
\newblock ISBN 9780511813658.
\newblock \doi{10.1017/CBO9780511813658}.

\bibitem[Woodroofe(1979)]{Woodroofe1979}
M.~Woodroofe.
\newblock A one-armed bandit problem with a concomitant variable.
\newblock \emph{Journal of the American Statistical Association}, 74\penalty0
  (368):\penalty0 799--806, 1979.
\newblock \doi{10.1080/01621459.1979.10481033}.

\end{thebibliography}

\appendix	

\section{Proof of Main Result}
\subsection{Proof of Theorem \ref{theorem_Perturbed-regret-bound}} \label{appendix_Perturbed-theorem}

We now present a proof of Theorem \ref{theorem_Perturbed-regret-bound}. The main results required are Lemma \ref{lemma_Perturbed-order_r}, Proposition \ref{proposition_Perturbed-Beta-bound} and Proposition \ref{proposition_Perturbed-lambdamin-bound} as stated in the body of the paper. We also require two more standard lemmas, Lemma \ref{Lem3} and Lemma \ref{lemma_Perturbed-alpha_sum}, which are stated and proved immediately after the proof of Theorem \ref{theorem_Perturbed-regret-bound}.

\Beginproof[Proof of Theorem \ref{theorem_Perturbed-regret-bound}.]
Using Lemma \ref{lemma_Perturbed-order_r} and Lemma \ref{Lem3}, we can derive  the regret over selling horizon $T$ in terms of $\left\|\hat{\bm{\beta}}_{t} - \bm{\beta}_{0}\right\|^2$, that is
\begin{align}
	\mathcal{R} \! g(T)
&
=\sum_{t=1}^{T} r_t\left(\bm{x}_{t}, {y}_{t} \,; \bm{\beta}_{0}\right) - r_t\left(\bm{x}^\star_{t}, {y}^\star_{t} \,; \bm{\beta}_{0}\right) 
\notag \\
    & =  \sum_{t=1}^{T} r\left(\bm{p}_{t}, \bm{c}_{t} \,; \bm{\beta}_{0}\right) - r\left(\p^{\star}(\bm{c}_{t}), \bm{c}_{t} \,; \bm{\beta}_{0}\right) 
\notag \\
&+
\sum_{t=1}^{T} r_t\left(\bm{x}_{t}, {y}_{t} \,; \bm{\beta}_{0}\right) - r\left(\bm{p}_{t}, \bm{c}_{t} \,; \bm{\beta}_{0}\right) 
+
\sum_{t=1}^{T} r\left(\bm{p}_{t}^{\star}(\bm{c}_{t}^{\star}), \bm{c}_{t} \,; \bm{\beta}_{0}\right) - r_t\left(\bm{x}_{t}^{\star}, {y}_{t}^{\star} \,; \bm{\beta}_{0}\right)  \notag \\
    & \leq  \sum_{t=1}^{T} \left|r\left(\bm{p}_{t}, \bm{c}_{t} \,; \bm{\beta}_{0}\right) - r\left(\p^{\star}(\bm{c}_{t}), \bm{c}_{t} \,; \bm{\beta}_{0}\right)\right| 
    \notag \\
    & \quad + 8 \sqrt{2 r_{\max} T \log T}
    \notag \\
    & \leq 8 \sqrt{2 r_{\max} T \log T} + K_{0}  \sum_{t=1}^{T}\left\|\bm{p}_{t}-\p^{\star}(\bm{c}_{t})\right\|^2  
    \notag  \\ 
    & \leq 8 \sqrt{2 r_{\max} T \log T}+ K_{0} \sum_{t=1}^{T}  
    2 \left\|\bm{p}_{t}-\bm{p}^{\star}(\bm{c}_{t}, \hat{\bm{\beta}}_{t})\right\|^2 
    +  K_{0} \sum_{t=1}^{T}  
    2 \left\|\bm{p}^{\star}(\bm{c}_{t}, \hat{\bm{\beta}}_{t})-\p^{\star}(\bm{c}_{t})\right\|^2  
    \notag \\ 
    & \leq  8 \sqrt{2 r_{\max} T \log T}+
    2 K_{0} u_{\max}^{2}  \sum_{t=1}^{T}\alpha_{t}^{2} 
    +  
    2 K_{0} K_{1} \sum_{t=1}^{T} \left\|\hat{\bm{\beta}}_{t} - \bm{\beta}_{0}\right\|^2 \,.
    \label{eq_Perturbed-regret_1}
 \end{align}
The first inequality above is immediate from Lemma \ref{Lem3}.
The second inequality applies \eqref{eq_Perturbed-r_bound_p} proven in Lemma \ref{lemma_Perturbed-order_r}.
The third inequality follows $(a+b)^2 \leq 2 a^2 + 2 b^2$ for any $a, b \in \R$.
The last inequality follows by the definition of the perturb price in \eqref{eq_Perturbed-perturb_p} and also from Lemma \ref{lemma_Perturbed-order_r}.

Applying Proposition \ref{proposition_Perturbed-Beta-bound}, we obtain, for some constant $K_2$,
\begin{equation*}\label{eq_Perturbed-beta_order}
    \left\|\hat{\bm{\beta}}_{t} - {\bm{\beta}}_{0}\right\|^{2}
    \leq K_2 \frac{\log (t)}{\lambda_{\min} (t)}\,.
\end{equation*}
By Proposition \ref{proposition_Perturbed-lambdamin-bound}, we have, for some $K_{3}$ and some the constant $T_{0}>0$, for any $t \geq T_{0}$,
\begin{align}\label{eq_Perturbed-lambda_simplified}
	\lambda_{\min} \left(t\right)
	\geq K_{3} \sum_{s=1}^{t}\alpha_{s}^{2} \,.
\end{align}
Applying \eqref{eq_Perturbed-lambda_simplified} to \eqref{eq_Perturbed-regret_1} gives that   
\begin{align*}
    \begin{split}
    \mathcal{R} \! g(T)
    \leq
8 \sqrt{2 r_{\max} T \log T}+
    2 K_{0} u_{\max}^{2}  \sum_{t=1}^{T}\alpha_{t}^{2} 
    +  
    2 K_{0} K_{1} \beta_{\max} T_{0}
    +
    \frac{2 K_{0} K_{1} K_2}{K_{3}}
    \sum_{t=1}^{T}\frac{\log (t)}{\sum_{s=1}^{t}\alpha_{s}^{2}}\,,
    \end{split}
\end{align*}
Thus from the form of the above, we can see that there exist constants $A_0, A_1$ and $A_2$, such that 
\begin{align*}
    \begin{split}
    \mathcal{R} \! g(T)
    \leq
    A_0 \sqrt{T\log T}
    +  
    A_1 \sum_{t=1}^{T}\alpha_{t}^{2} 
    +
    A_2  \sum_{t=2}^{T}\frac{\log(t)}{\sum_{s=1}^{t}\alpha_{s}^{2}}\,.
    \end{split}
\end{align*}
Finally notice that we take $\alpha_t = t^{-\eta}$ for $\eta \in [1/4, 1/2)$.

By a simple calculation shown in Lemma \ref{lemma_Perturbed-alpha_sum}, we can derive a specific upper bound on regret,
\begin{align*}
    \begin{split}
    \mathcal{R} \! g(T)
    & \leq
    A_0 \sqrt{T\log T}
    +  
    A_1 \frac{T^{1-2\eta} - \eta}{ 1-2\eta}
    +
    A_2 \left({1-2\eta}\right) \sum_{t=2}^{T}\frac{\log(t)}{t^{1-2\eta} - 1} \\
    & =
    \mathcal{O} \left(\sqrt{T\log T}+ T^{1-2\eta} + T^{2 \eta}\log(T)\right)\,.
    \end{split}
\end{align*}
Choosing $\eta = \frac{1}{4}$ gives
\begin{align*}
    \begin{split}
        \mathcal{R} \! g(T)
        & =   \mathcal{O} \left(\sqrt{T} \log(T)\right)\,.
    \end{split}
\end{align*}
Thus, we obtain the upper-bound on the regret. 
\Endproof

We now cover the additional lemmas required above.
Lemma \ref{Lem3} applied above is an application of the Azuma-Hoeffding Inequality. 

\begin{lemma}\label{Lem3}
With probability $1$, for any sequence $\bm x_t$, it eventually holds that
\begin{align*}
  \left| \sum_{t=1}^T r_t(\bm x_t , y_t ; \bm \beta_0) -r(\bm x_t ; \bm \beta_0)  \right|
\leq 4 \sqrt{2 r_{\max} T \log (T)}
\end{align*}
\end{lemma}
\Beginproof
Since $r(\bm x_t ; \bm \beta_0)$ is the conditional distribution of $r_t(\bm x_t , y_t ; \bm \beta_0)$,	 the summands of 
\begin{align*}
  M_T = \sum_{t=1}^T r_t(\bm x_t , y_t ; \bm \beta_0) -r(\bm x_t ; \bm \beta_0)
\end{align*}
form a martingale difference sequence. Thus by the Azzuma--Hoeffding Inequality (see \cite[E14.2]{Williams1991}) with $z_t =4 \sqrt{2 r_{\max}^2 t \log t}$
\begin{align*}
  \mathbb P \big( |M_t| \geq z_t \big)
\leq 
2 e^{- \frac{z_t^2}{8 t r_{\max}^2}}
=
\frac{2}{t^4}\, .
\end{align*}
Thus 
\begin{align*}
  \sum_{t=1}^\infty  \mathbb P ( | M_t | \geq z_t ) \leq \sum_{t=1}^\infty \frac{2}{t^4} <\infty \, .
\end{align*}
Thus by the Borel-Cantelli Lemma (see \cite[S2.7]{Williams1991}), with probability $1$, it holds that eventually $| M_t | \leq z_t$, which gives the result.
\Endproof

Lemma \ref{lemma_Perturbed-alpha_sum} is a standard integral test result.

\begin{lemma}\label{lemma_Perturbed-alpha_sum}
If we let $\alpha_t = t^{-\gamma}$ for $\gamma>0$, we have  
\[
\frac{t^{1-\gamma} - 1}{ 1-\gamma }  
\leq \sum_{s=1}^t  {s^{-\gamma}} 
\leq
1 + \frac{t^{1-\gamma} - 1}{ 1-\gamma }  \, .
\] 
\end{lemma}

\Beginproof 
	This can be obtained by simple calculations.
	Since $s^{-\gamma}$ is decreasing for $\gamma >0$, then
	\begin{align*}
		\sum_{s=1}^t  {s^{-\gamma}} \leq 1 + \int_{1}^{t}{s^{-\gamma}} \df s = 1 + \frac{t^{1-\gamma} - 1}{ 1-\gamma } \,,
	\end{align*}
	and
	\begin{align*}
		\sum_{s=1}^t  {s^{-\gamma}} \geq \int_{1}^{t}{s^{-\gamma}} \df s = \frac{t^{1-\gamma} - 1}{ 1-\gamma } \,.
	\end{align*}
Thus, we obtain the result.
\Endproof


\section{Proofs of Additional Results}\label{appendix_Perturbed-additionalresults}

In this section, we prove the supporting results Lemma \ref{lemma_Perturbed-order_r}, Proposition \ref{proposition_Perturbed-Beta-bound} and Proposition \ref{proposition_Perturbed-lambdamin-bound}. 
Lemma \ref{lemma_Perturbed-order_r} is an application of the Implicit Function Theorem.
Proposition 1 develops of the results of \cite{LaiWei1982} to the case of GLMs and follows similar lines to \cite{ChenHY1999} and \citet{Li2017}. 
The proof of Proposition \ref{proposition_Perturbed-lambdamin-bound} is a little more involved and requires an additional eigenvalue bound Proposition \ref{proposition_Perturbed-lambdaM}, which is proven using a Schur Decomposition. We can then use Proposition \ref{proposition_Perturbed-lambdaM}  along with random matrix theory bounds from \citet{Vershynin2018} to prove Proposition 2. 

We now define some notations that will be used throughout this section.

\paragraph{Additional Notation.}
For a positive definite matrix $A \in \mathbb{R}^{d \times d}$, we have $\|\bm{x}\|_{A}=\sqrt{\bm{x}^{\top} A \bm{x}}$. 
We denote by $V_{t} = \sum_{s=1}^{t} \bm{x}_{s}\bm{x}_{s}^\top$.
For $\rho>0$ we denote the closed ball around $\bm \beta_0$ and its boundary by  
\[
B_{\rho}(\bm \beta_0) := \left\{{\bm{\beta}} : \left\|{{\bm{\beta}}}-{\bm{\beta}}_{0}\right\| \leq \rho\right\}\qquad 
\text{and} \qquad \partial B_{\rho}(\bm \beta_0) = \left\{{\bm{\beta}} : \left\|{{\bm{\beta}}}-{\bm{\beta}}_{0}\right\|= \rho \right\}\,.
\]  
Let $a \wedge b = \min \{a,b\}$ and $a \vee b = \max \{a,b\}$ for $a,b\in \mathbb{R}$.
For a $d \times d$ symmetric matrix $M$, we define the operator norm (or spectral norm) by
\begin{align*}
	\|M\|_{op} 
	= \max_{\substack{x \in \mathbb{R}^{d} \\ x\neq 0}} \frac{\|M \bm{x}\|}{\|\bm{x}\|}
	=\max_{\substack{\bm{x} \in S^{d-1}}}\|M \bm{x}\|	=\max_{\substack{\bm{x} , \bm{y} \in S^{d-1}}} \bm{x}^{\top}M \bm{y}\,,
\end{align*}
where $S^{d-1}$ is the unit sphere in $\mathbb{R}^{d}$.
 We recall by the spectral theorem, a symmetric matrix has real valued eigenvalues and that its eigenvectors can be chosen as an orthogonal basis of $\mathbb{R}^{d}$.
For a square matrix $M$, we let $\lambda_{\min }(M)$ and $\lambda_{\max}(M)$ denote the minimum and maximum eigenvalues of $M$.
We use notations $M \succ 0$ to say that $M$ is positive definite and $M \succeq 0$ to say that $M$ is positive semi-definite.
If $M$ is a positive semi-definite matrix then
\[
\lambda_{\max}(M) = \|M\|_{op}\,,
\quad
\lambda_{\min}(M) = \min_{\bm{w}: \|\bm{w}\| = 1}\bm{w}^{\top}M \bm{w}\,,
\]
and thus 
\begin{align}
\lambda_{\min}(a_{1}M_{1}+a_{2}M_{2}) &\geq a_{1} \lambda_{\min}(M_{1}) + a_{2} \lambda_{\min}(M_{2})\, , 
\notag  \\
\lambda_{\max}(a_{1}M_{1}+a_{2}M_{2}) &\leq a_{1} \lambda_{\max}(M_{1}) + a_{2} \lambda_{\max}(M_{2})\, ,\label{l:max}\\
\lambda_{\min}\begin{pmatrix}
M_{1} & 0 \\
0 & M_{2}
\end{pmatrix}&
\geq \lambda_{\min}\left(M_{1}\right){\, \wedge \,} \lambda_{\min}\left(M_{2}\right) \notag \,.
\end{align}
for $a_1, a_2 \in \mathbb{R}_{+}$ and $M_1, M_2$ positive definite matrices.

\subsection{Proof of Lemma \ref{lemma_Perturbed-order_r}.}
We now prove Lemma \ref{lemma_Perturbed-order_r} which requires the Implicit Function Theorem (Theorem 9.28 of \cite{rudin1976principles}).

\Beginproof[Proof of Lemma \ref{lemma_Perturbed-order_r}.]
We first prove \eqref{eq_Perturbed-r_bound_p}.
Since $\nabla_{\bm{p}}r\left(\bm{p}^{\star}(\c), \bm{c}\,; \bm{\beta}_{0}\right) = 0$. 
A Taylor expansion w.r.t. $\bm{p}$ gives 
	\[
	r\left(\bm{p}, \bm{c}\,; \bm{\beta}_0\right) - r \left(\p^{\star}, \bm{c}\,; \bm{\beta}_{0}\right)  
	\leq
	\frac{1}{2} \left(\bm{p}- \bm{p}^{\star}\right)^{\top}  \nabla^{2} r\left(\tilde{\bm{p}}, \bm{c}\,; \bm{\beta}_{0}\right)\left(\bm{p}- \bm{p}^{\star}\right)^{\top}\,,
	\]
	for $\tilde{\bm{p}}$ on the line segment between $\bm{p}$ and $\bm{p}^{*}$. 
	Thus	
	\[
    \left|r\left(\bm{p}, \bm{c}\,; \bm{\beta}_0\right) 
    - r \left(\p^{\star}, \bm{c}\,; \bm{\beta}_{0}\right)  \right|
    \leq \frac{1}{2}\max_{\bm{c}\in \mathcal{C}} \max_{\tilde{\bm{p}}\in \mathcal{P}}
    \left\|\nabla^{2}r\left(\tilde{\bm{p}}, \bm{c}\,; \bm{\beta}_{0}\right)\right\|_{op}
     \big\|\p-\bm{p}^{\star} \big\|^2 \, .
    \]
Define
\[K_{0} := \max_{\bm{c}\in \mathcal{C}} \max_{\tilde{\bm{p}}\in \mathcal{P}} 
	\left\|\nabla^{2}r\left(\tilde{\bm{p}}, \bm{c}\,; \bm{\beta}_{0}\right)\right\|_{op}\,.
	\]
Notice $ K_{0} < \infty$, since $r\left(\bm{p}, \bm{c}\,; \bm{\beta}_{0}\right)$ is twice continuously differentiable and the sets $\mathcal{C}$ and $\mathcal{P}$ are compact.
The result in \eqref{eq_Perturbed-r_bound_p} is proved.

We now apply the Implicit Function Theorem to bound $\left\|\p-\bm{p}^{\star}\right\|^2$. 
We first consider the case under Assumption \ref{ass_Perturbed-A_finite}, and then under Assumption \ref{ass_Perturbed-A_convex}.

Under Assumption \ref{ass_Perturbed-A_finite}, the Implicit Function Theorem implies that for each $\bm{c}\in \mathcal{C}$, there exists a neighborhood $V_{c} \subseteq \R^{d}$ such that $\bm{p}^{\star}\left(\bm{c}\,; \bm{\beta} \right)$ is uniquely defined and is continuously differentiable in $\mathcal{B}$.
Thus taking $V = \cap_{\bm{c}\in \mathcal{C}} V_{c} $ and applying the Taylor expansion, for all $\bm{\beta} \in V$ give
\begin{align}\label{A1r}
\left\|\bm{p}^{\star}(\bm{c}\,; \bm{\beta})-\bm{p}^{\star}(\c)\right\|
\leq k_1
\left\|{\bm{\beta}} - \bm{\beta}_{0}\right\|	
\end{align}
where $k_1:=\sup_{\bm{c}\in \mathcal{C}}\sup_{\bm{\beta} \in V} 
\left|\nabla_{\bm{\beta}}\,\bm{p}^{\star}(\bm{c}\,; \bm{\beta}) \right|$.
The constant $k_1$ is finite since $\nabla_{\bm{\beta}}\,\bm{p}^{\star}(\bm{c}\,; \bm{\beta})$ is continuous, $\mathcal{C}$ is finite and $V$ can be chosen to be contained in a compact set.

Under Assumption \ref{ass_Perturbed-A_convex}, we know by strict concavity that for each $\mathcal{B}$ and $\mathcal{C}$, $\bm{p}^{\star} \left(\bm{c}\,; \bm{\beta} \right)$ is unique.
Further by Assumption \ref{ass_Perturbed-A_convex}, $\nabla_{\bm{p}}^{2}\, 
	r\left(\bm{p}, \bm{c}\,; \bm{\beta}_{0}\right)$ is invertible. 
	So the Implicit Function Theorem applies to 
	$\nabla_{\bm{p}} r\left(\bm{p}, \bm{c}\,; \bm{\beta}_{0}\right)$.
For all $\bm{c}\in \mathcal{C}$, $\bm{\beta} \in \mathcal{B}$, again applying the Taylor expansion gives
	\[
	\left\|\bm{p}^{\star}(\bm{c}\,; \bm{\beta})-\bm{p}^{\star}(\c)\right\|
	\leq
	\sup_{\bm{c}\in \mathcal{C}}\sup_{\bm{\beta} \in \mathcal{B}} 
	\left\|\nabla_{\bm{\beta}}\,\bm{p}^{\star}(\bm{c}\,; \bm{\beta}) \right\|_{op}
	\left\|{\bm{\beta}} - \bm{\beta}_{0}\right\|
	\,.
	\]
	Also by the Implicit Function Theorem, $\bm{p}^{\star} = \bm{p}^{\star}(\bm{c}\,; \bm{\beta})$ satisfies,
	\[
	\nabla_{\bm{\beta}}\,\bm{p}^{\star} 
	= -\left[\nabla_{\bm{p}}^{2}\, 
	r\left(\bm{p}^{\star}, \bm{c}\,; \bm{\beta}\right)\right]^{-1}
	\nabla_{\bm{p}, \bm{\beta}} \, 
	r\left(\bm{p}^{\star}, \bm{c}\,; \bm{\beta}\right)\,.
	\] 
	Thus, by the definition of the operator norm
	\begin{align*}
		\left\|\nabla_{\bm{\beta}}\,\bm{p}^{\star} \right\|
		& \leq 
		\left\|\nabla_{\bm{p}}^{2}\, 
		r\left(\bm{p}^{\star}, \bm{c}\,; \bm{\beta}\right)^{-1}\right\|_{op}
		\left\|\nabla_{\bm{p}, \bm{\beta}} \, 
		r\left(\bm{p}^{\star}, \bm{c}\,; \bm{\beta}\right)\right\|_{op}\\
		& \leq 
		\alpha^{-1}	\sup_{\bm{p},\bm{c},\bm{\beta}} \left\|\nabla_{\bm{p}, \bm{\beta}}\, r\left(\bm{p}^{\star}, \bm{c}\,; \bm{\beta}\right)\right\|_{op}\,,
	\end{align*}
	where we use the fact that $r\left(\bm{p}, \bm{c}\,; \bm{\beta}\right)$ is twice continuously differentiable and 
	$\min_{\bm{p},\bm{c},\bm{\beta}}\left\|\nabla_{\bm{p}}^{2}\, 
		r\left(\bm{p}^{\star}, \bm{c}\,; \bm{\beta}\right)\right\|_{op} \geq \alpha$ since $r(\bm{p},\bm{c}\,; \bm{\beta})$ is $\alpha$-strongly concave.
	Further $k_2:=\sup_{\bm{p},\bm{c},\bm{\beta}} \left\|\nabla_{\bm{p}, \bm{\beta}} r\left(\bm{p}^{\star}, \bm{c}\,; \bm{\beta}\right)\right\|_{op} < \infty$ because $r$ is continuously differentiable and because of the compactness of $\mathcal P$, $\mathcal C$ and $\mathcal B$.
Thus
\begin{align}\label{A2r}
	\left\|\bm{p}^{\star}(\bm{c}\,; \bm{\beta})-\bm{p}^{\star}(\c)\right\|
	\leq
	K_1
	\left\|{\bm{\beta}} - \bm{\beta}_{0}\right\|
	\,
\end{align}
for $K_1= k_2/\alpha$.

We have both  \eqref{A1r} and \eqref{A2r} holding. In other-words, under both Assumptions \ref{ass_Perturbed-A_finite} and \ref{ass_Perturbed-A_convex}, we have that 
	\[
	\sup_{\c\in\mathcal{C}}\left\|\bm{p}^{\star}(\bm{c}\,; \bm{\beta})-\bm{p}^{\star}(\bm{c})\right\|
	\leq K_{1}
	\left\|{\bm{\beta}} - \bm{\beta}_{0}\right\|\,,
	\]
	for all $\bm{\beta} \in V$ in a neighborhood of $\bm \beta$.
\Endproof
	
\subsection{Proof of Proposition \ref{proposition_Perturbed-Beta-bound}}

Proposition \ref{proposition_Perturbed-Beta-bound} develops of the results of \cite{LaiWei1982} to the case of GLMs. The proof follows similar lines to \cite{ChenHY1999} and \citet{Li2017} and then applies Lemma 1iii) of \cite{LaiWei1982}. 
We note there are some reported errors in the proofs of \cite{ChenHY1999}. So like \cite{Li2017}, we must take some care to work around these issues and make sure the proof method is applicable in our setting.

\Beginproof[Proof of Proposition \ref{proposition_Perturbed-Beta-bound}.]
We define $Z_{t} = \sum_{s=1}^{t}\epsilon_{s}\bm{x}_{s}$, where $\epsilon_{s}=y_s - \mu({\bm \beta}_0^{\top} \bm x_s)$. Our proof proceeds by bounding $||Z_t||_{V_t^{-1}}$ above and below. 

The bound in Proposition \ref{proposition_Perturbed-Beta-bound} is trivial if $\lambda_{\min }(t) =0$. Thus we assume that the increasing function $\lambda_{\min}(t)$ is positive at $t$. 
We define 
$$G_{t}(\bm{\beta}) := \sum_{s=1}^{t}
    \bm{x}_{s}
    \left({\mu\left({\bm{\beta}}^{\top}\bm{x}_{s}\right)} - {\mu\left({\bm{\beta}}_{0}^{\top}\bm{x}_{s}\right)} \right)\, .$$
Clearly, 
    $G_{t}(\bm{\beta}_{0}) = 0$. 
    Further since $\hat{\bm \beta}_t$ is defined to solve
    \[
    \sum_{s=1}^t \mu(\hat{\bm \beta}_t \bm x_s ) \bm x_s =  \sum_{s=1}^t y_s \bm x_s\, .
    \]
Thus holds that $G_{t}(\hat{\bm{\beta}}_t) = Z_{t}$ because
\begin{align*}
  G_{t}(\hat{\bm{\beta}}_t) &= \sum_{s=1}^{t}
    \bm{x}_{s}
    {\mu\left(\hat {\bm{\beta}}_t^{\top}\bm{x}_{s}\right)} - \sum_{s=1}^{t}\bm{x}_{s} {\mu\left({\bm{\beta}}_{0}^{\top}\bm{x}_{s}\right)} 
\\
&
=\sum_{s=1}^t y_s \bm x_s - \sum_{s=1}^{t}\bm{x}_{s} {\mu\left({\bm{\beta}}_{0}^{\top}\bm{x}_{s}\right)} 
=  \sum_{s=1}^{t}\bm{x}_{s} \epsilon_{s}= Z_{t}\,.
\end{align*}
The function $\mu(\cdot)$ is continuously differentiable and strictly increasing. Thus
by the Mean Value Theorem, there exists $\tilde{\bm{\beta}}$ on the line segment between $\hat{\bm{\beta}}_{t}$ and $\bm{\beta}_{0}$ such that
\begin{align}
        G_{t}(\hat{\bm{\beta}}_{t}) - G_{t}(\bm{\beta}_{0}) 
        & =  
        \sum_{s=1}^{t}
        \bm{x}_{s}
        \left(
        	\mu\left(\hat{\bm{\beta}}_{t}^{\top}\bm{x}_{s}\right)
        -\mu\left(\bm{\beta}_{0}^{\top}\bm{x}_{s}\right) 
        \right)\notag \\
        & =          
        \sum_{s=1}^{t}
        \dot{\mu}\left(\tilde{\bm{\beta}}^{\top}\bm{x}_{s}\right)
        \bm{x}_{s}\bm{x}_{s}^{\top}
        \left(\hat{\bm{\beta}}_{t} - {\bm{\beta}}_{0}\right)  
        = \nabla G_{t}(\tilde{\bm{\beta}})\left(\hat{\bm{\beta}}_{t} - {\bm{\beta}}_{0}\right) \, .\label{G_tag}
\end{align}
where 
$
\nabla G_t(\bm \beta)$ is the derivative of $G_t$.

Given Assumption \ref{ass_Perturbed-A2}, we define
\[
\kappa= \min_{\substack{\bm x,\bm \beta: \|\bm x\|\leq  x_{\max},\\ \|\bm \beta - \bm \beta_0\|\leq  \beta_{\max}} } \dot{\mu}({ \bm \beta}^\top \bm x )\,.
\]
We know that, by assumption, $\kappa >0$.
Thus from \eqref{G_tag}, we know that 
\[
\nabla G_t(\tilde{\bm \beta}) \succeq \kappa V_t \succeq \kappa \lambda_{\min}(t) I \,,
\]
and thus $V^{-1}_t  \succeq \kappa \nabla G_t(\tilde{\bm \beta})^{-1} $.
This implies that 
\begin{align*}
    \begin{split}
        \|G_{t}(\hat{\bm{\beta}}_t)\|_{V_{t}^{-1}}^{2} 
        & = \|G_{t}(\hat{\bm{\beta}}_t) - G_{t}({\bm{\beta}}_{0})\|_{V_{t}^{-1}}^{2} \\
        & = \left(\hat{\bm{\beta}}_t-{\bm{\beta}_{0}}\right)^{\top} 
       \nabla G_t(\tilde{\bm \beta})
        V_{t}^{-1}
       \nabla G_t(\tilde{\bm \beta})
        \left(\hat{\bm{\beta}}_t-{\bm{\beta}_{0}}\right) 
        \\
        &
        \geq
        \kappa 
        \left(\hat{\bm{\beta}}_t-{\bm{\beta}_{0}}\right)^{\top}
        \nabla G_t(\tilde{\bm \beta})
        \left(\hat{\bm{\beta}}_t-{\bm{\beta}_{0}}\right) 
        \\
        & 
        \geq \kappa^{2} \lambda_{\min}(V_{t})\|\hat{\bm{\beta}}_t-{\bm{\beta}_{0}} \|^{2} \, .
    \end{split}
\end{align*}
Since $Z_{t} = {G}_{t}(\hat{\bm{\beta}}_{t})$, 
\begin{equation*}
    \left\|Z_t\right\|^{2}_{V_{t}^{-1}} \geq \kappa^{2} \lambda_{\min}(V_{t})\left\|{\hat{\bm{\beta}}_{t}}-{\bm{\beta}_{0}} \right\|^{2}\,.
\end{equation*}
{To obtain the upper bound on $\left\|Z_t\right\|^{2}_{V_{t}^{-1}}$, we use Lemma \ref{lemma_Perturbed-LaiWei} (stated below), that almost surely}
\begin{align*}
	\left\|Z_t\right\|^{2}_{V_{t}^{-1}} 
    & = 
    O \left(\log \lambda_{\max}(t)\right)\, .
\end{align*}
Combining the two bounds above yields
\[
||Z_t||_{V^{-1}_t}^2 = O\left(\frac{\log \lambda_{\max}(t)}{\lambda_{\min}(t)}\right)\, .
\]
Finally with the observation that $\lambda_{\max}(t) \leq t x_{\max}^2$ (which follows from \eqref{l:max}), we then obtain the result.
\Endproof

The following lemma is a restatement of Lemma 1iii) in \cite{LaiWei1982}. For its proof, we refer to \cite{LaiWei1982}.
\begin{lemma}[\cite{LaiWei1982}]\label{lemma_Perturbed-LaiWei}
	For 
	\[
	Q_t = 
	Z^\top_t V^{-1}_t Z_t\,,
	\]
	on the event $\{\lim_{t \to \infty}\lambda_{\max}(t)=\infty\}$ it holds that almost surely
	\[
	Q_t = O \left(\log \lambda_{\max}(t)\right)\, .
	\]
\end{lemma}


\subsection{Proof of Proposition \ref{proposition_Perturbed-lambdamin-bound} }
To prove Proposition \ref{proposition_Perturbed-lambdamin-bound}, we derive a new eigenvalue bound that is critical to our proof.
This algebraic eigenvalue bound is given in Proposition \ref{proposition_Perturbed-lambdaM}.
\begin{proposition}\label{proposition_Perturbed-lambdaM}
For any symmetric matrix of the form 
$M=\begin{pmatrix}
A & B \\
B^{\top} & C
\end{pmatrix}$, 
we have
\[
\lambda_{\min}\left(M\right)
\geq 
\frac{\lambda_{\min}\left(C\right)^{2}}{\left(\|B\|_{op}+\lambda_{\min}\left(C\right)\right)^{2}+\lambda_{\min}\left(C\right)^{2}}
\left\{\lambda_{\min}\left(A-B C^{-1} B^{\top}\right)\wedge \lambda_{\min}\left(C\right)\right\}\,.
\]
\end{proposition}

To prove Proposition \ref{proposition_Perturbed-lambdamin-bound}, we apply Proposition \ref{proposition_Perturbed-lambdaM} along with a random matrix bound from \cite{Vershynin2018}. We state and prove this result (as Lemma \ref{lemma_Perturbed-boundnorm}) after the proof of Proposition \ref{proposition_Perturbed-lambdamin-bound}.
Also we require Lemma \ref{lemma_Perturbed-Approx_isometries} which is a standard eigenvalue bound stated and the concentration bound Lemma \ref{lemma_Perturbed-boundnorm} after the proof of Proposition \ref{proposition_Perturbed-lambdamin-bound}.  
Now, we give the proof of Proposition \ref{proposition_Perturbed-lambdamin-bound}. 
\Beginproof[Proof of Proposition \ref{proposition_Perturbed-lambdamin-bound}.]
 	Applying the shorthand $\bm{p}_{s} = \bm{p}_{ce}\left(\bm{c}_s, \hat{\bm{\beta}}_{s}\right)$.
	We expand the design matrix as follows
	\begin{align*}
		\sum_{s=1}^{t} \bm{x}_{s} \bm{x}_{s}^\top 
		& = \sum_{s=1}^{t} \left(\bm{p}_{s}+\alpha_{s} \bm{z}_{s}, \bm{c}_s \right)\left(\bm{p}_{s}+\alpha_{s} \bm{z}_{s}, \bm{c}_s \right)^\top \\
		& = \sum_{s=1}^{t} \left(\bm{p}_{s}, \bm{c}_s \right)\left(\bm{p}_{s}, \bm{c}_s \right)^\top 
		+ \alpha_{s}^{2}\bm{z}_{s}\bm{z}_{s}^\top 
		+ \alpha_{s} \bm{z}_{s} \left(\bm{p}_{s}, \bm{c}_s \right)^\top 
		+ \alpha_{s}\left(\bm{p}_{s}, \bm{c}_s \right)\bm{z}_{s}^\top \,.
	\end{align*}
	By Lemma \ref{lemma_Perturbed-Approx_isometries}, we have 
	\begin{align}\label{eq_Perturbed-lambda_xx}
		\begin{split}
			\lambda_{\min} \left( \sum_{s=1}^{t} \bm{x}_{s} \bm{x}_{s}^\top \right)
			& \geq
			\lambda_{\min} \left(\sum_{s=1}^{t} \left(\bm{p}_{s}, \bm{c}_s \right)\left(\bm{p}_{s}, \bm{c}_s \right)^\top 
			+ \alpha_{s}^{2}\bm{z}_{s}\bm{z}_{s}^\top\right)
			- \left\|\sum_{s=1}^{t} \alpha_{s} \bm{z}_{s} \left(\bm{p}_{s}, \bm{c}_s \right)^\top 
			+ \alpha_{s}\left(\bm{p}_{s}, \bm{c}_s \right)\bm{z}_{s}^\top\right\|_{op}
			\,.
		\end{split}
	\end{align}
	By Proposition \ref{proposition_Perturbed-lambdaM} we can bound the first term in \eqref{eq_Perturbed-lambda_xx}
	\begin{align}\label{eq_Perturbed-lambda_xz}
		\begin{split}
			\lambda_{\min} \left(\sum_{s=1}^{t} \left(\bm{p}_{s}, \bm{c}_s \right)\left(\bm{p}_{s}, \bm{c}_s \right)^\top 
			+ \alpha_{s}^{2}\bm{z}_{s}\bm{z}_{s}^\top\right)
			& \geq
			\frac{\lambda_{\min}\left(\sum_{s=1}^{t} \bm{c}_{s} \bm{c}_{s}^\top\right)^{2}}{\left(\left\|\sum_{s=1}^{t} \bm{p}_{s} \bm{c}_s^{\top}\right\|_{op}+\lambda_{\min}\left(\sum_{s=1}^{t} \bm{c}_{s} \bm{c}_{s}^\top\right)\right)^{2}
			+\lambda_{\min}\left(\sum_{s=1}^{t} \bm{c}_{s} \bm{c}_{s}^\top\right)^{2}}\\
			& \times \left\{\lambda_{\min}\left(\sum_{s=1}^{t}\bm{p}_{s}\bm{p}_{s}^\top + \alpha_{s}^{2}\bm{z}_{s}\bm{z}_{s}^\top\right)
			\wedge \lambda_{\min}\left(\sum_{s=1}^{t} \bm{c}_{s} \bm{c}_{s}^\top\right)\right\}\,.
		\end{split}
	\end{align}
	We now provide lower-bounds on the various terms given above. 
	First, since sets $\mathcal{P}$ and $\mathcal{C}$ are bounded, we have 
	\begin{align}\label{eq_Perturbed-bound_pc}
		\left\|\sum_{s=1}^{t} \bm{p}_{s} \bm{c}_s^{\top}\right\|_{op}
		\leq t \,{p}_{\max} \, c_{\max}\,.
	\end{align}
	Second, by Lemma \ref{lemma_Perturbed-Approx_isometries} and Lemma \ref{lemma_Perturbed-boundnorm}, we have
	\begin{align}\label{eq_Perturbed-lambda_SigmaC}
		\begin{split}
			\lambda_{\min}\left( \sum_{s=1}^{t} \bm{c}_{s} \bm{c}_{s}^{\top}\right)
			& \geq t \lambda_{\min} \left( \Sigma^{c} \right) 
			- \left\|\sum_{s=1}^{t} \left(\bm{c}_{s} \bm{c}_{s}^\top - \Sigma^{c}\right) \right\|_{op} \,,
		\end{split}
	\end{align}
	where
	\begin{align*}
		\begin{split}
			\left\|\sum_{s=1}^{t} \left(\bm{c}_{s} \bm{c}_{s}^\top - \Sigma^{c}\right) \right\|_{op} 
			\leq
			 \sqrt{16 t \log(t)  \left({c}_{\max}^2 + \left\|\Sigma^{c}\right\|_{op}\right)^2}\,,
		\end{split}
	\end{align*}
	and also 
	\begin{align}\label{eq_Perturbed-lambda_SigmaC2}
		\begin{split}
		\lambda_{\min}\left( \sum_{s=1}^{t} \bm{c}_{s} \bm{c}_{s}^{\top}\right) \leq \lambda_{\max}\left( \sum_{s=1}^{t} \bm{c}_{s} \bm{c}_{s}^{\top}\right) \leq t c_{\max}^2\,.
		\end{split}
	\end{align} 
	Similarly, for the last term, by Lemmas \ref{lemma_Perturbed-Approx_isometries} and \ref{lemma_Perturbed-boundnorm},
	\begin{align}\label{eq_Perturbed-lambda_pz}
			\lambda_{\min}\left(\sum_{s=1}^{t}\bm{p}_{s}\bm{p}_{s}^\top + \alpha_{s}^{2}\bm{z}_{s}\bm{z}_{s}^\top\right)
			& \geq 
			\lambda_{\min}\left( \alpha_{s}^{2}\bm{z}_{s}\bm{z}_{s}^\top\right)
			\notag \\
			& \geq 
			\sum_{s=1}^t \alpha_{s}^2 \lambda_{\min} (\Sigma^z) - \left\| \sum_{s=1}^t\alpha_{s}^{2}\left(\bm{z}_{s}\bm{z}_{s}^{\top}-{\Sigma^z}\right)\right\|_{op}
			\notag \\
			& \geq 
			\sum_{s=1}^t \alpha_{s}^2 \lambda_{\min} (\Sigma^z) 
			-  
			\sqrt{16 \log(t)  \left({z}_{\max}^2 + \left\|\Sigma^{z}\right\|_{op}\right)^2\sum_{s=1}^{t}\alpha_{s}^{4}}\,.
	\end{align}
	The first inequality is obtained since $\sum_{s=1}^{t}\bm{p}_{s}\bm{p}_{s}^\top$ is positive definite matrix and 
	\[
	\lambda_{\min}\left(\sum_{s=1}^{t}\bm{p}_{s}\bm{p}_{s}^\top \right) >0 \,.
	\]
	Applying \eqref{eq_Perturbed-bound_pc},  \eqref{eq_Perturbed-lambda_SigmaC}, \eqref{eq_Perturbed-lambda_SigmaC2} and \eqref{eq_Perturbed-lambda_pz} to \eqref{eq_Perturbed-lambda_xz} gives the result 
	\begin{align*}
		\begin{split}
			& \lambda_{\min} \left(\sum_{s=1}^{t} \left(\bm{p}_{s}, \bm{c}_s \right)\left(\bm{p}_{s}, \bm{c}_s \right)^\top 
			+ \alpha_{s}^{2}\bm{z}_{s}\bm{z}_{s}^\top\right)\\
			& \geq \frac{\left(\lambda_{\min}\left(\Sigma^{c}\right) - \sqrt{{\log(t)}/{t}}\right)^{2}}{\left(p_{\max}c_{\max}+ c_{\max}^{2}\right)^{2} + c_{\max}^{2}}\\
			& \times 
			\left\{\lambda_{\min}\left({\Sigma^z}\right)\sum_{s=1}^{t}\alpha_{s}^{2} - \sqrt{16 \log(t)  \left({z}_{\max}^2 + \left\|\Sigma^{z}\right\|_{op}\right)^2\sum_{s=1}^{t}\alpha_{s}^{4}}\right\}
			\wedge
			\left\{t \lambda_{\min}\left(\Sigma^{c}\right)- \sqrt{16 t \log(t)  \left({c}_{\max}^2 + \left\|\Sigma^{c}\right\|_{op}\right)^2}\right\}\,.
		\end{split}
	\end{align*}
	We can simplify it as following.  
There exists a constant time $T_{0}$ depending on $\Sigma^{c}$ and $\Sigma^{z}$ such that
	\begin{align*}
		\begin{split}
			& \left\{\lambda_{\min}\left({\Sigma^z}\right)\sum_{s=1}^{t}\alpha_{s}^{2} - \sqrt{16 \log(t)  \left({z}_{\max}^2 + \left\|\Sigma^{z}\right\|_{op}\right)^2\sum_{s=1}^{t}\alpha_{s}^{4}}\right\}
			\wedge
			\left\{t \lambda_{\min}\left(\Sigma^{c}\right)- \sqrt{16 t \log(t)  \left({c}_{\max}^2 + \left\|\Sigma^{c}\right\|_{op}\right)^2} \right\} \\
			& \geq \sum_{s=1}^{t}\alpha_{s}^{2} \frac{\lambda_{\min}\left(\Sigma^{z}\right)}{2}\, ,
		\end{split}
	\end{align*}
	for all $t\geq T_0$.

Now for the 2nd term in \eqref{eq_Perturbed-lambda_xx}.
	By triangle inequality, we have
	\begin{align*}
	\left\|\sum_{s=1}^{t} \alpha_{s} \bm{z}_{s} \left(\bm{p}_{s}, \bm{c}_s \right)^\top 
	+ \alpha_{s}\left(\bm{p}_{s}, \bm{c}_s \right)\bm{z}_{s}^\top\right\|_{op}
	& \leq 
	\left\|\sum_{s=1}^{t} \alpha_{s} \bm{z}_{s} \left(\bm{p}_{s}, \bm{c}_s \right)^\top \right\|_{op} 
	+ \left\|\sum_{s=1}^{t} \alpha_{s}\left(\bm{p}_{s}, \bm{c}_s \right)\bm{z}_{s}^\top\right\|_{op} \\
	& \leq 
	2 \left\|\sum_{s=1}^{t} \alpha_{s} \bm{z}_{s} \left(\bm{p}_{s}, \bm{c}_s \right)^\top \right\|_{op}\,.
	\end{align*}
	Using Lemma \ref{lemma_Perturbed-boundnorm_xz}, eventually it holds that,
	\[\left\|\sum_{s=1}^{t} \alpha_{s} \bm{z}_{s} \left(\bm{p}_{s}, \bm{c}_s \right)^\top \right\|_{op} 
	\leq \sqrt{16 \left({z}_{\max}{x}_{\max}\right)^2 \log(t)  \sum_{s=1}^{t}\alpha_{s}^{2}}\,.
	\]
	Combining the above two inequalities, we obtain that for sufficiently large $t$,
	\begin{align*}
		\begin{split}
			\lambda_{\min} \left( \sum_{s=1}^{t} \bm{x}_{s} \bm{x}_{s}^\top \right)
			\geq \sum_{s=1}^{t}\alpha_{s}^{2} \frac{\lambda_{\min}\left(\Sigma^{z}\right)}{2}
-
 \sqrt{16 \log(t)  \left({z}_{\max}{x}_{\max}\right)^2\sum_{s=1}^{t}\alpha_{s}^{2}}
			\,,
		\end{split}
	\end{align*}
Notice that, for $\alpha_t \propto {1}/{t^\eta}$ for $\eta < 1/2$, the term
$\sqrt{\log (t)\sum_{s=1}^{t}\alpha_{s}^{2}}$, above, 
is dominated by $\sum_{s=1}^{t}\alpha_{s}^{2}$, and thus, we obtain the required result.
\Endproof

The following is a well-known eigenvalue bound. 
\begin{lemma}
\label{lemma_Perturbed-Approx_isometries}
Let $A, B$ be symmetric positive definite matrices of size $d \times d$.
Then, 
\begin{equation*}
    \lambda_{\min}\left(A\right) \geq \lambda_{\min}\left(B\right) - \left|\left| A - B \right|\right|_{op} \,.
\end{equation*}
\end{lemma}
\Beginproof 
Recall that 
\[\lambda_{\min}\left(A\right)=\min_{\bm{x}: \|\bm{x}\|=1} \bm{x}^{\top}A \bm{x}.\]
We can write
\begin{align*}
    \begin{split}
        \bm{x}^{\top}A \bm{x} = \bm{x}^{\top}(A-B) \bm{x} + \bm{x}^{\top}B \bm{x} \,.
    \end{split}
\end{align*}
For the first term, by Cauchy-Schwartz inequality we have 
\[
-\bm{x}^{\top}(A-B) \bm{x} 
\leq|\langle \bm{x},(A-B) \bm{x} \rangle| 
\leq\|\bm{x}\|\|(A-B) \bm{x}\| 
\leq\|A-B\|_{op} \,.
\]
For the second term, we know
\[\bm{x}^{\top} B \bm{x} \geq \lambda_{\min }(B)\,.\]
Then we have
\[
\bm{x}^{\top} A \bm{x}=\bm{x}^{\top}(A-B) \bm{x}+\bm{x}^{\top} B \bm{x} \geq - \|A-B\|_{op} + \lambda_{\min }(B)\,.
\]
Therefore,
\[
\lambda_{\min }(A) \geq \lambda_{\min }(B)-\|A-B\|_{op}
\,.\]
We obtain the result.
\Endproof

The following result gives a concentration bound on covariance matrices. The proof can be found in \citet{Vershynin2018}. (Also see \cite{vershynin2012close}.)

\begin{lemma}[\citet{Vershynin2018}]
\label{lemma_Perturbed-boundnorm}
If $\bm{x}_{s} \in \mathbb{R}^{d}$ are bounded such that, for all $s\geq 1$  
\[
\mathbb{E}\left[ \bm{x}_{s}\mid\mathcal{F}_{s-1}\right]=\bm{0}
\qquad \text{and} \qquad 
\mathbb{E}\left[\bm{x}_{s} \bm{x}_{s}^{\top}\mid\mathcal{F}_{s-1}\right] = \Sigma_{s}^{x} \,.
\]
We assume that there exists ${x}_{\max} \in \mathbb{R}^{+}$, such that bound $\| \bm{x}_{s} \|_{\infty} \leq {x}_{\max} $ with probability $1$ and $\Sigma_{s}^{x}$ is positive definite. 
Then, 
\[
\mathbb{P}\left(\left\|\sum_{s=1}^{t}\left(\bm{x}_{s} \bm{x}_{s}^{\top} - \Sigma_{s}^{x}  \right) \right\|_{op} \geq \varepsilon \right)
\leq 
2 \cdot 9^{2d} \exp\left\{-\frac{\left({\varepsilon}/{2}\right)^{2}}{2 \sum_{s=1}^{t} \left({x}_{\max}^{2} + \left\|\Sigma_{s}^{x} \right\|_{op}\right)^2}\right\}\,.
\]
\end{lemma}
\Beginproof 
	We show this argument in two steps.
	We first control $\sum_{t=1}^{T}\bm{x}_t \bm{x}^\top_t$ over 
	a $\varepsilon$-net, and then extend the bound to the full supremum norm by a continuity argument.
	
	Using Lemma \ref{lemma_Perturbed-number_Euclideanball} (stated below) and choosing $\varepsilon = \frac{1}{4}$ and, we can find an $\varepsilon$-net $\mathcal{N}$ of the unit sphere $S^{d-1}$ with cardinality 
	\begin{equation*}
		\left|\mathcal{N}\right| \leq 9^{d} \,.
	\end{equation*}
	By Lemma \ref{lemma_Perturbed-Quadratic_net} (stated below), the operator norm of $\bm{x}_s \bm{x}^\top_s$ can be bounded on $\mathcal{N}$, that is
	\begin{align}\label{eq_Perturbed-operatornorm}
        \left\|\sum_{s=1}^{t}\left(\bm{x}_{s}\bm{x}_{s}^{\top}-\Sigma_{s}^{x}  \right)\right\|_{op} 
        & \leq   
        2 \max_{\bm{v}, \bm{w}\in \mathcal{N}} 
        \left\langle \left(\sum_{s=1}^{t}\left(\bm{x}_{s}\bm{x}_{s}^{\top}- \mathbb{E}\left[ \bm{x}_{s} \bm {x}_{s}^{\top} \right]\right)\right) \bm{v}, \bm{w} \right\rangle \notag \\
        & \leq  
        2 \max_{\bm{v}, \bm{w}\in \mathcal{N}} 
        \left|\bm{v}^{\top}\left(\sum_{s=1}^{t}\left(\bm{x}_{s}\bm{x}_{s}^{\top}- \mathbb{E}\left[ \bm{x}_{s} \bm {x}_{s}^{\top} \right]\right)\right)\bm{w} \right| \,.
	\end{align}
	We first fix $\bm{v}, \bm{w} \in \mathcal{N}$, and by Azuma--Hoeffding inequality\footnote{We note that Vershynin applies a Hoeffding bound. This is the only substantive difference in the proof here.}
for any $\varepsilon > 0$ we can state that 
	\begin{align*}
	\begin{split}
        \mathbb{P}\left(\left|\bm{v}^{\top}\left(\sum_{s=1}^{t}\left(\bm{x}_{s}\bm{x}_{s}^{\top}- \mathbb{E}\left[ \bm{x}_{s} \bm {x}_{s}^{\top} \mid\mathcal{F}_{s-1}\right]\right)\right)\bm{w} \right|  \geq \frac{\varepsilon}{2} \right) 
        \leq 
        2 \, \exp \left\{-\frac{ (\varepsilon/2)^{2}}{2 \sum_{s=1}^{t} \left({x}_{\max}^{2} + \left\|\Sigma_{s}^{x} \right\|_{op}\right)^2}\right\}  \,.
    \end{split}
	\end{align*}
	Given that $\| \bm{x}_{s} \|_{\infty} \leq {x}_{\max} $, we have $\left\|\bm{x}_{s}\bm{x}_{s}^{\top}- \mathbb{E}\left[ \bm{x}_{s} \bm {x}_{s}^{\top}\mid\mathcal{F}_{s-1} \right]\right\| \leq {x}_{\max}^{2} + \left\|\Sigma_{s}^{x} \right\|_{op}$.
	
	Next, we unfix $\bm{v}, \bm{w} \in \mathcal{N}$ using a union bound. 
	Since that $\mathcal{N}$ has cardinality bounded by $9^{d}$, we obtain
	\begin{align*}
    \begin{split}
        &\quad \mathbb{P}\left(\max_{\bm{v}, \bm{w}\in \mathcal{N}} 
        \left|\bm{v}^{\top}\left(\sum_{s=1}^{t}\left(\bm{x}_{s}\bm{x}_{s}^{\top}- \mathbb{E}\left[ \bm{x}_{s} \bm {x}_{s}^{\top} \mid\mathcal{F}_{s-1}\right]\right)\right)\bm{w} \right| \geq  \frac{\varepsilon}{2} \right)         
\\
& \leq 
        \sum_{\bm{v}, \bm{w}\in \mathcal{N}} \mathbb{P}\left( 
        \left|\bm{v}^{\top}\left(\sum_{s=1}^{t}\left(\bm{x}_{s}\bm{x}_{s}^{\top}- \mathbb{E}\left[ \bm{x}_{s} \bm {x}_{s}^{\top} \mid\mathcal{F}_{s-1}\right]\right)\right)\bm{w} \right| \geq  \frac{\varepsilon}{2}\right) \\
        & \leq 
        |\mathcal{N}|^2 \cdot 2 \, \exp \left\{-\frac{ (\varepsilon/2)^{2}}{2\sum_{s=1}^{t} \left({x}_{\max}^{2} + \left\|\Sigma_{s}^{x} \right\|_{op}\right)^2}\right\}\\
        & \leq 
        9^{2d} \cdot 2 \, \exp \left\{-\frac{ \varepsilon^{2}}{8\sum_{s=1}^{t} \left({x}_{\max}^{2} + \left\|\Sigma_{s}^{x} \right\|_{op}\right)^2}\right\}\,.
    \end{split}
	\end{align*}
	Together with \eqref{eq_Perturbed-operatornorm}, we have
	\begin{align*}
    \begin{split}
        \mathbb{P}\left(\left\|\sum_{s=1}^{t}\left(\bm{x}_{s}\bm{x}_{s}^{\top}-\Sigma_{s}^{x} \right)\right\|_{op}  \geq \varepsilon \right) 
        & = 
        \mathbb{P}\left(2 \max_{\bm{v}, \bm{w}\in \mathcal{N}} 
        \left|\bm{v}^{\top}\left(\sum_{s=1}^{t}\left(\bm{x}_{s}\bm{x}_{s}^{\top}- \mathbb{E}\left[ \bm{x}_{s} \bm {x}_{s}^{\top} \mid\mathcal{F}_{s-1}\right]\right)\right)\bm{w} \right|  \geq \varepsilon \right) \\
        & \leq 
        2 \cdot 9^{2d} \exp \left\{-\frac{ \varepsilon ^{2}}{8\sum_{s=1}^{t} \left({x}_{\max}^{2} + \left\|\Sigma_{s}^{x} \right\|_{op}\right)^2}\right\}
        \,.
    \end{split}
	\end{align*}
Thus, we obtain the result.
\Endproof
A straight-forward consequence of this result is the following.
\begin{corollary}\label{corollary_Perturbed-boundnorm}
	With probability $1$, eventually in $t$ it holds that
	\[
	\left\| \sum_{s=1}^{t}\left(\bm{x}_{s}\bm{x}_{s}^{\top}-\Sigma_{s}^{x}  \right) \right\|_{op}
	\leq
	\sqrt{16 \log(t) \sum_{s=1}^{t} \left({x}_{\max}^{2} + \left\|\Sigma_{s}^{x} \right\|_{op}\right)^2}\,.
	\]
\end{corollary}
\Beginproof[Proof of Corollary \ref{corollary_Perturbed-boundnorm}.]
	Notice, if we set 
	\[
	 \varepsilon_{t} = \sqrt{16 \log(t) \sum_{s=1}^{t} \left({x}_{\max}^{2} + \left\|\Sigma_{s}^{x} \right\|_{op}\right)^2}\,,
	\]
	then,
	\[
	\mathbb{P}\left(\left\| \sum_{s=1}^{t}\left(\bm{x}_{s}\bm{x}_{s}^{\top}-\Sigma_{s}^{x}\right)\right\|_{op} \geq \varepsilon_{t}  \right)
	\leq 
	\frac{2 \cdot 9^{2d} }{t^2}\,.
	\]
	By the Borel--Cantelli Lemma, the result holds. 
\Endproof

\begin{lemma}
\label{lemma_Perturbed-boundnorm_xz} 
\[
\mathbb{P}\left(\left\|\sum_{s=1}^{t} \alpha_s \bm{z}_s \bm{x}_{s}^{\top} \right\|_{op} \geq \varepsilon \right)
\leq 
2 \cdot 9^{2d} \exp\left\{-\frac{\left({\varepsilon}/{2}\right)^{2}}{2 \left({z}_{\max}{x}_{\max}\right)^2 \sum_{s=1}^{t} \alpha_s^2}\right\}\,.
\]
and thus, with probability $1$, eventually it holds that
\begin{align}\label{eq:evs}
  \left\|\sum_{s=1}^{t} \alpha_s \bm{z}_s \bm{x}_{s}^{\top} \right\|_{op}
\leq 
	\sqrt{ 16 z^2_{\max}x^2_{\max}\log(t) \sum_{s=1}^t \alpha^2  }
\end{align}
\end{lemma}

\Beginproof
Similar to the proof of Lemma \ref{lemma_Perturbed-boundnorm}, we show this argument in two steps.
	We first control $\sum_{s=1}^{t}\alpha_s \bm{z}_s \bm{x}_{s}^{\top}$ over 
	a $\varepsilon$-net, and then extend the bound to the full supremum norm by a continuity argument.
Notice that the summands of $\sum_{s=1}^{t}\alpha_s \bm{z}_s \bm{x}_{s}^{\top}$ are a bounded martingale difference sequence.
	
	Using Lemma \ref{lemma_Perturbed-number_Euclideanball} (stated below) and choosing $\varepsilon = \frac{1}{4}$ and, we can find an $\varepsilon$-net $\mathcal{N}$ of the unit sphere $S^{d-1}$ with cardinality 
	\begin{equation*}
		\left|\mathcal{N}\right| \leq 9^{d} \,.
	\end{equation*}
	By Lemma \ref{lemma_Perturbed-Quadratic_net} (stated below), the operator norm can be bounded by terms on $\mathcal{N}$, that is
	\begin{align}\label{eq_Perturbed-operatornormxz}
        \left\|\sum_{s=1}^{t}\alpha_s \bm{z}_s \bm{x}_{s}^{\top}\right\|_{op} 
        & \leq   
        2 \max_{\bm{v}, \bm{w}\in \mathcal{N}} 
        \left\langle \left(\sum_{s=1}^{t}\alpha_s \bm{z}_s \bm{x}_{s}^{\top}\right) \bm{v}, \bm{w} \right\rangle \notag \\
        & \leq  
        2 \max_{\bm{v}, \bm{w}\in \mathcal{N}} 
        \left|\bm{v}^{\top}\left(\sum_{s=1}^{t}\alpha_s \bm{z}_s \bm{x}_{s}^{\top}\right) \bm{w} \right| \,.
	\end{align}
	We first fix $\bm{v}, \bm{w} \in \mathcal{N}$, and by Azuma--Hoeffding inequality, for any $\varepsilon > 0$ we can state that 
	\begin{align*}
	\begin{split}
        \mathbb{P}\left(\left|\bm{v}^{\top}\left(\sum_{s=1}^{t}\alpha_s \bm{z}_s \bm{x}_{s}^{\top}\right) \bm{w} \right|  \geq \frac{\varepsilon}{2} \right) 
        \leq 
        2 \, \exp \left\{-\frac{ (\varepsilon/2)^{2}}{2 \sum_{s=1}^{t} \left(\alpha_s {z}_{\max}{x}_{\max}\right)^2}\right\}  \,.
    \end{split}
	\end{align*}
	Next, we unfix $\bm{v}, \bm{w} \in \mathcal{N}$ using a union bound. 
	Since that $\mathcal{N}$ has cardinality bounded by $9^{d}$, we obtain
	\begin{align*}
    \begin{split}
        \mathbb{P}\left(\max_{\bm{v}, \bm{w}\in \mathcal{N}} 
        \left|\bm{v}^{\top}\left(\sum_{s=1}^{t}\alpha_s \bm{z}_s \bm{x}_{s}^{\top}\right) \bm{w} \right| \geq  \frac{\varepsilon}{2} \right) 
        & \leq 
        \sum_{\bm{v}, \bm{w}\in \mathcal{N}} \mathbb{P}\left( 
        \left|\bm{v}^{\top}\left(\sum_{s=1}^{t}\alpha_s \bm{z}_s \bm{x}_{s}^{\top}\right) \bm{w} \right| \geq  \frac{\varepsilon}{2}\right) \\
        & \leq 
        |\mathcal{N}|^2 \cdot 2 \, \exp \left\{-\frac{ (\varepsilon/2)^{2}}{2\sum_{s=1}^{t} \left(\alpha_s {z}_{\max}{x}_{\max}\right)^2}\right\}\\
        & \leq 
        9^{2d} \cdot 2 \, \exp \left\{-\frac{ \varepsilon^{2}}{8\sum_{s=1}^{t} \left(\alpha_s {z}_{\max}{x}_{\max}\right)^2}\right\}\,.
    \end{split}
	\end{align*}
	Together with \eqref{eq_Perturbed-operatornormxz}, we have
	\begin{align*}
    \begin{split}
        \mathbb{P}\left(\left\|\sum_{s=1}^{t} \alpha_s \bm{z}_s \bm{x}_{s}^{\top} \right\|_{op} \geq \varepsilon \right)
        & = 
        \mathbb{P}\left(2 \max_{\bm{v}, \bm{w}\in \mathcal{N}} 
        \left|\bm{v}^{\top}\left(\sum_{s=1}^{t}\alpha_s \bm{z}_s \bm{x}_{s}^{\top}\right) \bm{w} \right|  \geq \varepsilon \right) \\
        & \leq 
        2 \cdot 9^{2d} \exp \left\{-\frac{ \varepsilon^{2}}{8\left({z}_{\max}{x}_{\max}\right)^2\sum_{s=1}^{t} \alpha_s^2}\right\}
        \,.
    \end{split}
	\end{align*}
Thus, we obtain the result. For \eqref{eq:evs}, the argument follows in an identical manner to Corollary \ref{corollary_Perturbed-boundnorm}.\Endproof

Lemmas~\ref{lemma_Perturbed-number_Euclideanball} and \ref{lemma_Perturbed-Quadratic_net} are stated below.
For the proofs of Lemmas~\ref{lemma_Perturbed-number_Euclideanball} and \ref{lemma_Perturbed-Quadratic_net}, we refer to Section 4 in \citet{Vershynin2018}.  

\begin{lemma}[Covering Numbers of the Euclidean ball]
\label{lemma_Perturbed-number_Euclideanball}
The covering numbers of the unit Euclidean ball is such that, for any $\varepsilon>0$
\[
\left(\frac{1}{\varepsilon}\right)^{d} 
\leq \mathcal{N}\leq\left(\frac{2}{\varepsilon}+1\right)^{d} \,.
\]
The same upper bound is true for the unit Euclidean sphere $S^{d-1}$.	
\end{lemma}
\begin{lemma}[Quadratic Form on a Net]\label{lemma_Perturbed-Quadratic_net}
	Let $A$ be an $m \times n$ matrix and $\varepsilon \in[0,1 / 2)$.
	For any $\varepsilon$-net $\mathcal{N}$ of the sphere $S^{n-1}$ and any $\varepsilon$-net $\mathcal{M}$ of the sphere $S^{m-1}$, we have
\begin{equation}\label{eq_Perturbed-Quadratic_net}
	\sup_{x \in \mathcal{N}, y \in \mathcal{M}}\langle A x, y\rangle \leq\|A\|_{op} \leq \frac{1}{1-2 \varepsilon} \sup_{x \in \mathcal{N}, y \in \mathcal{M}}\langle A x, y\rangle \,.
\end{equation}
Moreover, if $m=n$ and $A$ is symmetric, then 
\[
\sup _{x \in \mathcal{N}}|\langle A x, x\rangle| \leq\|A\|_{op}  \leq \frac{1}{1-2 \varepsilon}  \sup _{x \in \mathcal{N}}|\langle A x, x\rangle|\,.
\]
\end{lemma}




\subsection{Proof of Proposition \ref{proposition_Perturbed-lambdaM}.}
\label{proof_proposition_Perturbed-lambdaM}

Proposition \ref{proposition_Perturbed-lambdaM} gives a new eigenvalue bound. It essentially allows us to isolate the randomness caused by contextual information. We can then subsequently apply random matrix concentration bounds to this. Proposition  \ref{proposition_Perturbed-lambdaM} requires two lemmas, Lemma \ref{lemma_Perturbed-M1} which is a standard result on the Schur Complement, and Lemma \ref{lemma_Perturbed-bound_f(p)} which is a straightforward calculus argument. These are stated and proven immediately after the proof of Proposition \ref{proposition_Perturbed-lambdaM}.

\Beginproof[Proof of Proposition \ref{proposition_Perturbed-lambdaM}.]
We want to lower-bound the minimum eigenvalue of $M$, where $M$ is a positive semi-definite matrix, and thus have a bound for the ``only if" direction of Lemma \ref{lemma_Perturbed-M1}.
We show that a positive semi-definite matrix containing a positive definite sub-matrix can be made positive definite by adding an appropriate sub-matrix.

Applying Lemma \ref{lemma_Perturbed-M1} gives
\begin{align}\label{eq_Perturbed-lambda_M}
	\lambda_{\min}\left(M\right) 
	& = \min_{\bm{w}:\|\bm{w}\| = 1} \bm{w}^{\top}M\bm{w}  \notag \\
	& = \min_{\bm{w}:\|\bm{w}\| = 1} \bm{w}^{\top}\begin{pmatrix}
	I & BC^{-1} \notag \\
	0 & I
	\end{pmatrix}
	\begin{pmatrix}
	A-BC^{-1}B^{\top} & 0  \notag \\
	0 & C
	\end{pmatrix}
	\begin{pmatrix}
	I & BC^{-1} \\
	0 & I
	\end{pmatrix}^{\top} \bm{w}\\
	& = \min_{\bm{w}:\|\bm{w}\| = 1} 
		\left(\bm{w}_1, \bm{w}_1+BC^{-1}\bm{w}_2\right)\begin{pmatrix}
		A-BC^{-1}B^{\top} & 0 \\
		0 & C
		\end{pmatrix}
		\begin{pmatrix}
		\bm{w}_1 \\
		\bm{w}_1+BC^{-1}\bm{w}_2
		\end{pmatrix} \,.
\end{align}
Given the vector applied above, we now lower bound of the vector $\left(\bm{w}_1, \bm{w}_1+BC^{-1}\bm{w}_2\right)$.
\begin{align*}
\begin{split}
	\left\|
	\left(\bm{w}_1, \bm{w}_2+BC^{-1}\bm{w}_1\right)
	\right\|^{2}
	& = \left\|\bm{w}_1
	\right\|^{2} + \left\|\bm{w}_2+BC^{-1}\bm{w}_1\right\|^{2}\\
	& \geq \left\|\bm{w}_1
	\right\|^{2} + \left(\left(\left\|\bm{w}_2\right\| - \left\|BC^{-1}\bm{w}_1\right\|\right) \vee 0 \right)^{2}\\
	& \geq \left\|\bm{w}_1
	\right\|^{2} + \left(\left(\left\|\bm{w}_2\right\| - \left\|BC^{-1}\right\|_{op}  \left\|\bm{w}_1\right\|\right) \vee 0 \right)^{2}\,.
\end{split}
\end{align*}
Since $\|\bm{w}\|^2 =\|\bm{w}_1\|^2 + \|\bm{w}_2\|^2 = 1$, we can write the above bound in the form of 
	\[f(p) = p + \left(\left(\sqrt{1-p} - b\sqrt{p}\right)\vee 0 \right)^{2}\,,
	\]
	where $p = \|\bm{w}_1\|^2$ and $b= \left\|BC^{-1}\right\|_{op}$.
	By Lemma \ref{lemma_Perturbed-bound_f(p)}, we have 
	\begin{equation}\label{eq_Perturbed-bound_wBC2}
		\left\|
		\left(\bm{w}_1, \bm{w}_1+BC^{-1}\bm{w}_2\right)
		\right\|^{2}
		\geq \frac{1}{ \left(\left\|BC^{-1}\right\|_{op} +1\right)^2 + 1 }\,.
	\end{equation}
	For any $c \geq 0$, 
	\begin{align}\label{eq_Perturbed-bound_vABC}
		\min_{\bm{v}:\|\bm{v}\| = c} 
		\bm{v}^{\top}
		\begin{pmatrix}
		A-BC^{-1}B^{\top} & 0 \\
		0 & C
		\end{pmatrix}
		\bm{v} 
		= c \lambda_{\min}\left(A-BC^{-1}B^{\top} \right) \wedge \lambda_{\min}\left(C\right) \,.
	\end{align}
	Thus, applying \eqref{eq_Perturbed-bound_wBC2} and \eqref{eq_Perturbed-bound_vABC} to \eqref{eq_Perturbed-lambda_M} gives
	\begin{align*}
		\begin{split}
			\lambda_{\min}\left(M\right) 
			& = \min_{\bm{w}:\|\bm{w}\| = 1} 
			\left(\bm{w}_1, \bm{w}_1+BC^{-1}\bm{w}_2\right)\begin{pmatrix}
			A-BC^{-1}B^{\top} & 0 \\
			0 & C
			\end{pmatrix}
			\begin{pmatrix}
			\bm{w}_1 \\
			\bm{w}_1+BC^{-1}\bm{w}_2
			\end{pmatrix} \\
			& \geq \min_{\bm{v}:\|\bm{v}\|\geq c} \bm{v}^{\top}
			\begin{pmatrix}
			A-BC^{-1}B^{\top} & 0 \\
			0 & C
			\end{pmatrix}\bm{v}\\
			& \geq \frac{\lambda_{\min}\left(A-BC^{-1}B^{\top} \right) \wedge \lambda_{\min}\left(C\right)}{ \left(\left\|BC^{-1}\right\|_{op}+1\right)^2 + 1 }\,.
		\end{split}
	\end{align*}
	Substituting,
	\[
	\left\|BC^{-1}\right\|_{op} = \frac{\left\|B\right\|_{op}}{\lambda_{\min}\left(C\right)}\,,
	\]
	gives the required result. 
\Endproof
The following is a well known lemma for matrices under a Schur decomposition.
\begin{lemma}\label{lemma_Perturbed-M1}
For any symmetric matrix $M$ of the form 
\begin{equation*}\label{eq_Perturbed-M}
M=
\begin{pmatrix}
A & B \\
B^{\top} & C
\end{pmatrix}\,,
\end{equation*}
if $C$ is invertible then $M \succ 0$ if and only if $C \succ 0$ and $A-B C^{-1} B^{\top} \succ 0$.
\end{lemma}
\Beginproof 
The Schur Complement of $C$ in $M$ is given by
\[
A-B C^{-1} B^{\top}\,,
\]
obviously the Schur Complement is symmetric.
Notice that we can use this to decompose $M$ as
\[
M=
\begin{pmatrix}
I & BC^{-1} \\
0 & I
\end{pmatrix}
\begin{pmatrix}
A-BC^{-1}B^{\top} & 0 \\
0 & C
\end{pmatrix}
\begin{pmatrix}
I & BC^{-1} \\
0 & I
\end{pmatrix}^{\top} \,.
\]
A block diagonal matrix is positive definite if and only if  each diagonal block is positive definite, which concludes the proof.
\Endproof

\begin{lemma}\label{lemma_Perturbed-bound_f(p)}
	For a function $f(p) = p + \left(\left(\sqrt{1-p} - b\sqrt{p}\right)\vee 0 \right)^{2}$, we have
	\begin{equation*}
		\min_{p\in[0,1]}f(p)\geq \frac{1}{ \left(b+1\right)^2 + 1 }\,.
	\end{equation*}
\end{lemma}
\Beginproof 
	Since $\sqrt{1-p} \geq 1- \sqrt{p}$, and then
	\[
	f(p) = p + \left(\left(1 - (b+1)\sqrt{p}\right)\vee 0 \right)^{2}\,.
	\]
	We define 
	\[
	g(x) = x^2 + \left(\left(1 - (b+1)x\right)\vee 0 \right)^{2}\,.
	\]
	Note the above bound states
	\[
	f(p)\geq g(\sqrt{p})\,.
	\]
	To obtain the result, we prove the claim that 
	\[
	g(x) \geq \frac{1}{ \left(b+1\right)^2 + 1 }\,.
	\]
	We can obtain this by finding the minimum of $g(x)$ and showing that $x^{\star}$ is the minimum value such that $1-(1+b)x > 0$.
	Note that function $g(\cdot)$ is convex with $g(0)=g(1)=1$, so finding a local minimum is sufficient.
	Now, we let $g'(x) = 0$, that is
	\begin{align*}
	 	\frac{\df}{\df x}\left\{ x^2 + \left(1 - (b+1)x \right)^{2}\right\} 
	 	= \frac{\df}{\df x}\left\{ ((b+1)^2+1)x^2 - 2\left((b+1)x \right)+1\right\} =0\,.
	\end{align*}
	This implies 
	\[
	x^{\star} = \frac{b+1}{(b+1)^2+1}\,,
	\]
	and 
	\[
	(x^{\star})^2 + \left(1 - (b+1)x^{\star} \right)^{2}= \frac{1}{(b+1)^2+1}\,.
	\]
	Further notice that $1-(1+b)x^{\star} > 0$, so this point is a minimum of the function $g(x)$.
	Thus, we have
	\[
	\min_{p\in[0,1]}f(p)\geq \min_{x\in[0,1]}g(x)\geq \frac{1}{ \left(b+1\right)^2 + 1 }\,.
	\] 
	The result is obtained.
\Endproof

\end{document}